\documentclass[11pt,ams]{article}
\usepackage{graphicx}
\usepackage{amssymb}
\usepackage{amsmath}
\newcommand{\MSbar}{\overline{\mbox{MS}}}

\newcommand{\p}{\partial}
\newcommand{\oc}{\overline{c}}

\newcommand{\omu}{\overline{\mu}}
\newcommand{\lms}{\Lambda_{\overline{\mbox{\tiny{MS}}}}}
\newcommand{\Evac}{E_{\mathrm{{vac}}}}
\newcommand{\mf}{\mathcal{F}}

\newcommand{\ol}{\overline{\lambda}}
\newcommand{\og}{\overline{g}}
\newcommand{\om}{\overline{m}}
\newcommand{\wrho}{\widehat{\varrho}}
\newcommand{\wm}{\widehat{m}}
\newcommand{\wl}{\widehat{\lambda}}

\newcommand{\sect}[1]{ \section{#1} \setcounter{equation}{0} }

\begin{document}
\date{}
\title{\textbf{The Gribov parameter and the dimension two gluon condensate in Euclidean Yang-Mills theories in the Landau gauge}}
\author{\textbf{D. Dudal}$^{a}$\thanks{%
Research Assistant of the Research Foundation - Flanders (FWO-Vlaanderen).} \ , \textbf{R.F. Sobreiro}$^{b}$\thanks{%
sobreiro@uerj.br}  \ , \textbf{S.P. Sorella}$^{b}$\thanks{%
sorella@uerj.br}{\ }\footnote{Work supported by FAPERJ, Funda{\c
c}{\~a}o de Amparo {\`a} Pesquisa do Estado do Rio de Janeiro,
under the program {\it Cientista do Nosso Estado},
E-26/151.947/2004.}  \ , \textbf{H. Verschelde}$^{a}$\thanks{%
david.dudal@ugent.be, henri.verschelde@ugent.be} \\\\
\textit{$^{a}$ Ghent University} \\
\textit{Department of Mathematical Physics and Astronomy} \\
\textit{Krijgslaan 281-S9, B-9000 Gent, Belgium}\\
[3mm] \textit{$^{b}$ UERJ, Universidade do Estado do Rio de Janeiro} \\
\textit{Rua S{\~a}o Francisco Xavier 524, 20550-013 Maracan{\~a}} \\
\textit{Rio de Janeiro, Brasil}} \maketitle

\begin{abstract}
\noindent The local composite operator $A_\mu^2$ is added to the
Zwanziger action, which implements the restriction to the Gribov
region $\Omega$ in Euclidean Yang-Mills theories in the Landau
gauge. We prove that Zwanziger's action with the inclusion of the
operator $A_\mu^2$ is renormalizable to all orders of perturbation
theory, obeying the renormalization group equations. This allows to
study the dimension two gluon condensate $\left\langle
A_\mu^2\right\rangle$ by the local composite operator formalism when
the restriction to the Gribov region $\Omega$ is taken into account.
The resulting effective action is evaluated at one-loop order in the
$\MSbar$ scheme. We obtain explicit values for the Gribov parameter
and for the mass parameter due to $\left\langle
A_\mu^2\right\rangle$, but the expansion parameter turns out to be
rather large. Furthermore, an optimization of the perturbative
expansion in order to reduce the dependence on the renormalization
scheme is performed. The properties of the vacuum energy, with or
without the inclusion of the condensate $\left\langle
A_\mu^2\right\rangle$, are investigated. In particular, it is shown
that in the original Gribov-Zwanziger formulation, i.e. without the
inclusion of the operator $A_\mu^2$, the resulting vacuum energy is
always positive at one-loop order, independently from the choice of
the renormalization scheme and scale. In the presence of
$\left\langle A_\mu^2\right\rangle$, we are unable to come to a
definite conclusion at the order considered. In the $\MSbar$ scheme,
we still find a positive vacuum energy, again with a relatively
large expansion parameter, but there are renormalization schemes in
which the vacuum energy is negative, albeit the dependence on the
scheme itself appears to be strong. Concerning the behaviour of the
gluon and ghost propagators, we recover the well known consequences
of the restriction to the Gribov region, and this in the presence of
$\left\langle A_\mu^2\right\rangle$, i.e. an infrared suppression of
the gluon propagator and an enhancement of the ghost propagator.
Such a behaviour is in qualitative agreement with the results
obtained from the studies of the Schwinger-Dyson equations and from
lattice simulations.
\end{abstract}

\vfill\newpage\ \makeatother

\sect{Introduction.} The dimension two condensate $\left\langle
A_\mu^2\right\rangle$ has received a great deal of attention in the
last few years, see for example
\cite{Gubarev:2000nz,Gubarev:2000eu,Verschelde:2001ia,Kondo:2001tm,Kondo:2001nq,Boucaud:2001st,Boucaud:2000nd,RuizArriola:2004en,Boucaud:2005rm,
Dudal:2003vv,Dudal:2002pq,Browne:2003uv,Browne:2004mk,Gracey:2004bk,Li:2004zu,Shakin:2004te,Gubarev:2005it}.
This condensate was already introduced in \cite{Lavelle:1988eg} in
order to analyze the gluon propagator within the Operator Product
Expansion (OPE), while in \cite{Greensite:1985vq} the condensate
$\left\langle A_i^2\right\rangle$ was considered in the Coulomb
gauge. A renormalizable effective potential for $\left\langle
A_\mu^2\right\rangle$ has been constructed and evaluated in analytic
form up to two-loop order in the Landau gauge within the local
composite operator (LCO) formalism in
\cite{Verschelde:2001ia,Browne:2003uv}. The output of these
investigations is that a non-vanishing condensate is favoured as it
lowers the vacuum energy. The renormalizability of the local
composite operator formalism, see \cite{Knecht:2001cc} for an
introduction to the method, was proven to all orders of perturbation
theory, in the case of $\left\langle A_\mu^2\right\rangle$, in
\cite{Dudal:2002pq} using the algebraic renormalization technique
\cite{Piguet:1995er}. Besides the Landau gauge, the method was
extended to other gauges as, for instance, the Curci-Ferrari gauge
\cite{Dudal:2003gu,Dudal:2003pe}, the linear covariant gauges
\cite{Dudal:2003np,Dudal:2003by} and, more recently, the maximal
Abelian gauge \cite{Dudal:2004rx}.
\\\\As a consequence of the existence of a non-vanishing
condensate $\left\langle A_\mu^2\right\rangle$, a dynamical mass
parameter for the gluons can be generated in the gauge fixed
Lagrangian, see \cite{Verschelde:2001ia,Browne:2003uv,Dudal:2003by}.
We mention that a gluon mass has been proven to be rather useful in
the phenomenological context, see e.g.
\cite{Parisi:1980jy,Field:2001iu,Giacosa:2004ug}. Moreover, mass
parameters are commonly used in the fitting formulas for the data
obtained in lattice simulations, where the gluon propagator has been
studied to a great extent in the Landau gauge
\cite{Cucchieri:1999sz,Bonnet:2001uh,Langfeld:2001cz,Cucchieri:2003di,Bloch:2003sk,Furui:2003jr,Furui:2004cx}.
\\\\The lattice results so far obtained have provided firm evidence of the
suppression of the gluon propagator in the infrared region, in the
Landau gauge. Next to the gluon propagator, also the ghost
propagator has been investigated numerically on the lattice
\cite{Bloch:2003sk,Furui:2003jr,Furui:2004cx,Suman:1995zg,Bakeev:2003rr},
exhibiting an infrared enhancement. It is worth remarking that, in
agreement with lattice results, this infrared behaviour of the gluon
as well as of the ghost propagator has been obtained in the analysis
of the Schwinger-Dyson equations, see
\cite{vonSmekal:1997is,vonSmekal:1997is2,Atkinson:1997tu,Atkinson:1998zc,Alkofer:2000wg,Watson:2001yv,Zwanziger:2001kw,Lerche:2002ep},
as well as in a study making use of the exact renormalizaton group
technique \cite{Pawlowski:2003hq}.
\\\\The aim of the present
work is to investigate further the condensation of the operator
$A_\mu^2$ in the Landau gauge using the local composite operator
formalism. This will be done by taking into account the
nonperturbative effects related to the existence of the Gribov
ambiguities \cite{Gribov:1977wm}, which are known to affect the
Landau gauge fixing condition, $\partial _{\mu }A_{\mu }^{a}=0$. As
a consequence of the existence of the Gribov copies, the domain of
integration in the path integral has to be restricted in a suitable
way. Gribov's orginal proposal was to restrict the domain of
integration to the region $\Omega$ whose boundary $\partial \Omega$
is the first Gribov horizon, where the first vanishing eigenvalue of
the Faddeev-Popov operator, $-\partial _{\mu }\left(
\partial _{\mu }\delta ^{ab}+gf^{acb}A_{\mu }^{c}\right) $, appears \cite{Gribov:1977wm}. Within
the region $\Omega $ the Faddeev-Popov operator is positive
definite, i.e. $-\partial _{\mu }\left( \partial _{\mu }\delta
^{ab}+gf^{acb}A_{\mu }^{c}\right) >0$. One of the main results of
Gribov's work \cite{Gribov:1977wm} was that the gluon, respectively
ghost propagator, got suppressed, respectively enhanced, in the
infrared due to the restriction to the region $\Omega$.
\\\\In two previous papers \cite{Sobreiro:2004us,Sobreiro:2004yj},
we have already worked out the consequences of the restriction to
the Gribov region $\Omega$ when the dynamical generation of a gluon
mass parameter due to $\left\langle A_\mu^2\right\rangle$ takes
place, also finding an infrared suppression of the gluon and an
enhancement of the ghost propagator. In \cite{Sobreiro:2004us}, we
closely followed the setup of Gribov's paper \cite{Gribov:1977wm}.
In this work, we shall rely on the Zwanziger local formulation of
the Gribov horizon. In a series of papers
\cite{Zwanziger:1989mf,Zwanziger:1992qr}, Zwanziger has been able to
implement the restriction to the Gribov region $\Omega$ through the
introduction of a nonlocal horizon function appearing in the
Boltzmann weight defining the Euclidean Yang-Mills measure. More
precisely, according to \cite{Zwanziger:1989mf,Zwanziger:1992qr},
the starting Yang-Mills measure in the Landau gauge is given by
\begin{equation}
 d\mu _{\gamma }=DA\delta (\partial_\mu A_\mu^a)\det
(\mathcal{M})e^{-\left( S_{\mathrm{YM}}+\gamma ^{4}H\right) }\;,
\label{m1}
\end{equation}
where
\begin{equation}
\mathcal{M}^{ab}=-\partial _{\mu }\left( \partial _{\mu }\delta
^{ab}+gf^{acb}A_{\mu }^{c}\right) \;,  \label{mm1}
\end{equation}
\begin{equation}
S_{\mathrm{YM}}=\frac{1}{4}\int d^{4}xF_{\mu \nu }^{a}F_{\mu \nu
}^{a}\;, \label{m2}
\end{equation}
and
\begin{equation}
H=\int d^{4}xh(x)=g^{2}\int d^{4}xf^{abc}A_{\mu }^{b}\left( \mathcal{M}%
^{-1}\right) ^{ad}f^{dec}A_{\mu }^{e}\;,  \label{m3}
\end{equation}
is the so-called horizon function, which implements the
restriction to the Gribov region. Notice that $H$ is nonlocal. The
parameter $\gamma$, known as the Gribov parameter, has the
dimension of a mass and is not free, being determined by the
horizon condition
\begin{equation}
\left\langle h(x)\right\rangle =4\left( N^{2}-1\right) \;,
\label{m4}
\end{equation}
where the expectation value $\left\langle h(x)\right\rangle $ has to
be evaluated with the measure $d\mu _{\gamma }$. To the first order,
the horizon condition (\ref{m4}) reads, in $d$ dimensions,
\begin{equation}
1=\frac{N\left( d-1\right) }{4}g^{2}\int \frac{d^{d}k}{\left( 2\pi
\right) ^{d}}\frac{1}{k^{4}+2Ng^{2}\gamma ^{4}}\;.  \label{mm5}
\end{equation}
This equation coincides with the original gap equation derived by
Gribov for the parameter $\gamma$ \cite{Gribov:1977wm}.
\\\\ Albeit
nonlocal, the horizon function $H$ can be localized through the
introduction of a suitable set of additional fields. As shown in
\cite{Zwanziger:1989mf,Zwanziger:1992qr,Maggiore:1993wq}, the
resulting local action turns out to be renormalizable to all orders
of perturbation theory. Remarkably, we shall be able to prove that
this feature is preserved when the local operator $A_\mu^2$ is
introduced in the Zwanziger action. Moreover, the resulting theory
turns out to obey a homogeneous renormalization group equation.
These important properties will allow us to study the condensation
of the operator $A_\mu^2$ within a local renormalizable framework
when the restriction to the Gribov region $\Omega$ is implemented.
\\\\It is worth remarking that the Gribov region is not free from
gauge copies
\cite{Semenov,Dell'Antonio:1991xt,Dell'Antonio2,vanBaal:1991zw},
i.e. Gribov copies still exist inside $\Omega $. To avoid the
presence of these additional copies, a further restriction to a
smaller region $\Lambda $, known as the fundamental modular region,
should be implemented. At present, a clear understanding of the role
played by these additional copies appears to be a very difficult
task. Nevertheless, we should mention that, recently, it has been
argued in \cite{Zwanziger:2003cf} that the additional copies
existing inside $\Omega$ could have no influence on the expectation
values, so that averages calculated over $\Lambda$ or $\Omega$ might
give the same value.\\\\The paper is organized as follows. In
section 2, we give a short account of how the nonlocal horizon
functional $H$ can be localized by means of the introduction of
additional fields. In section 3, we prove the renormalizability, to
all orders of perturbation theory, of Zwanziger's action in the
presence of the operator $A_\mu^2$, introduced through the local
composite operator formalism. As the model has a rich symmetry
structure, translated into several Ward identities, it turns out
that only three independent renormalization factors are necessary.
The resulting quantum effective action obeys a homogeneous
renormalization group equation, as explicitly verified at one-loop
order. From this effective action, two coupled gap equations,
associated to the condensate $\left\langle A_\mu^2\right\rangle$ and
to the Gribov parameter $\gamma$, are derived. Section 4 is devoted
to the study of these gap equations at one-loop order in the
$\MSbar$ renormalization scheme. It is worth mentioning that, under
certain conditions, we find that it is possible that the condensate
$\left\langle A_\mu^2\right\rangle$ is positive when the horizon
condition is imposed. We prove that in the $\MSbar$ scheme, and at
one-loop order, the solution of the gap equations is necessarily one
with $\left\langle A_\mu^2\right\rangle>0$.  We recall that without
the restriction to the Gribov region $\Omega$, the value found for
$\left\langle A_\mu^2\right\rangle$ using the local composite
operator formalism is negative, see
\cite{Verschelde:2001ia,Browne:2003uv,Dudal:2003by}. Let us also
mention here that in
\cite{Boucaud:2001st,Boucaud:2000nd,RuizArriola:2004en,Boucaud:2005rm},
a positive estimate for $\left\langle A_\mu^2\right\rangle$ was
obtained when using the OPE in combination with $\left\langle
A_\mu^2\right\rangle$. These works were based on the observation of
a certain discrepancy at relatively large momentum between the
expected perturbative behaviour and the obtained lattice behaviour
of e.g. the effective strong coupling constant and gluon propagator.
This discrepancy could be accounted for by power corrections in
$\frac{1}{q^2}$, due to a positive $\left\langle
A_\mu^2\right\rangle_{\mathrm{OPE}}$ gluon condensate. The presence
of such power corrections has also been discussed in
\cite{Furui:2005bu}. We do not know if there is a direct connection
between the condensate $\left\langle A_\mu^2\right\rangle$ that we
determine, and $\left\langle A_\mu^2\right\rangle_{\mathrm{OPE}}$,
as the latter is expected to contain only infrared contributions,
according to an OPE treatment, while the gap equations fixing the
gluon condensate and the Gribov parameter are evaluated using
perturbation theory, implying that reliable results are only to be
expected at a sufficiently large scale.\\\\Although the expansion
parameter proves to be rather large, an attempt to obtain explicit
values for the Gribov and gluon mass parameter is still presented.
Also, we shall prove that in the original Gribov-Zwanziger model,
the vacuum energy is always positive at one-loop order, irrespective
of the choice of renormalization scheme and scale. We outline the
importance of the sign of the vacuum energy, as it is related to the
gauge invariant gluon condensate $\left\langle
F_{\mu\nu}^2\right\rangle$, via the trace anomaly. From
\begin{equation}\label{traceano1}
\theta_{\mu\mu}=\frac{\beta(g^2)}{2g^2}F_{\mu\nu}^2\;,
\end{equation}
the vacuum energy can be traced back to the value of the gluon
condensate $\left\langle F_{\mu\nu}^2\right\rangle$. In particular,
for $N=3$, from this anomaly one deduces
\begin{equation}\label{traceano2}
\left\langle\frac{g^2}{4\pi^2}F_{\mu\nu}^2\right\rangle=-\frac{32}{11}\Evac\;,
\end{equation}
where the one-loop $\beta$-function has been used. Hence, a positive
vacuum energy implies a negative value for the condensate
$\left\langle\frac{g^2}{4\pi^2}F_{\mu\nu}^2\right\rangle$. This is
in contradiction with what is found. In QCD, with quarks present,
one can extract phenomenological values for
$\left\langle\frac{g^2}{4\pi^2}F_{\mu\nu}^2\right\rangle$ via the
sum rules \cite{Shifman:1978bx}, obtaining positive values for this
condensate. It was discussed in \cite{DiGiacomo:1998nu} how to
obtain an estimate for it by means of lattice calculations. In the
case of $N=3$ Yang-Mills theory without quarks, it was found that
\begin{equation}\label{traceano3}
\left\langle\frac{g^2}{4\pi^2}F_{\mu\nu}^2\right\rangle=0.14\pm0.02\mathrm{GeV}^4\;.
\end{equation}
Let us mention here that the Yang-Mills $\beta$-function is negative
up to the (known) four-loop order
\cite{vanRitbergen:1997va,Czakon:2004bu,Chetyrkin:2004mf}. Hence,
$\Evac$ and
$\left\langle\frac{g^2}{4\pi^2}F_{\mu\nu}^2\right\rangle$ will
continue to have opposite sign at higher order. From this viewpoint,
it seems to us that it would be an asset that the vacuum energy
obtained from any kind of calculation is at least negative.\\\\In
section 5 we present an optimized expansion in order to reduce the
dependence on the choice of renormalization scheme to a single
parameter $b_0$, related to the coupling constant renormalization.
This is achieved by exchanging the mass parameters by their
renormalization scale and scheme invariant counterparts and by
re-expanding the series in the one-loop coupling constant. For
$b_0=0$, which corresponds to the $\MSbar$ scheme, we find a
positive $\left\langle A_\mu^2\right\rangle$, positive $\Evac$ and
hence negative $\left\langle F_{\mu\nu}^2\right\rangle$. However, we
find that a region of $b_0$ is existing in which the vacuum energy
is negative, but unfortunately the dependence on $b_0$ in this
region happens to be very large. A higher order analysis seems to be
required to reach more definite conclusions about the sign of
$\left\langle A_\mu^2\right\rangle$, $\Evac$ or $\left\langle
F_{\mu\nu}^2\right\rangle$.
\\\\For the benefit of the reader, we provide in section 6 an overview
of some important consequences stemming from the presence of the
Gribov and gluon mass parameters on the gluon and ghost propagators.
We point out a particular renormalization property of the Zwanziger
action in order to ensure the enhancement of the ghost propagator.
Conclusions are written down in section 7, while the technical
details of our work have been collected in the Appendices A and B.

\sect{Local action from the restriction to the Gribov region.} As
explained in \cite{Zwanziger:1989mf,Zwanziger:1992qr}, the nonlocal
functional $H$ can be localized by means of the introduction of a
suitable set of additional ghost fields. More precisely, for the
localized version of the measure $d\mu _{\gamma }$ we get,
\begin{equation}
d\mu _{\gamma }=DADbDcD\overline{c}D\varphi D\overline{\varphi }D\omega D%
\overline{\omega }e^{-S}\;,  \label{l1}
\end{equation}
where $S$ is given by\footnote{%
Our conventions are different from those originally used by
Zwanziger. These can be obtained from ours by setting $\varphi
\rightarrow -\varphi $ and $\omega \rightarrow -\omega $.}
\begin{equation}
S=S_{0}-\gamma ^{2}g\int d^{4}x\left( f^{abc}A_{\mu }^{a}\varphi
_{\mu }^{bc}+f^{abc}A_{\mu }^{a}\overline{\varphi }_{\mu
}^{bc}\right)\;,  \label{ll1}
\end{equation}
while
\begin{eqnarray}
S_{0} =S_{\mathrm{YM}}&+&\int d^{4}x\;\left( b^{a}\partial_\mu A_\mu^{a}+\overline{c}%
^{a}\partial _{\mu }\left( D_{\mu }c\right) ^{a}\right) \;  \nonumber \\
&+&\int d^{4}x\left( \overline{\varphi }_{\mu }^{ac}\partial _{\nu
}\left(
\partial _{\nu }\varphi _{\mu }^{ac}+gf^{abm}A_{\nu }^{b}\varphi _{\mu
}^{mc}\right) -\overline{\omega }_{\mu }^{ac}\partial _{\nu }\left( \partial
_{\nu }\omega _{\mu }^{ac}+gf^{abm}A_{\nu }^{b}\omega _{\mu }^{mc}\right)
\right.  \nonumber \\
 &\;&\;\;\;\;\;\;\;\;\;\;\;-g\left.\left( \partial _{\nu }\overline{\omega }_{\mu
}^{ac}\right) f^{abm}\left( D_{\nu }c\right) ^{b}\varphi _{\mu
}^{mc}\right) \;.  \label{l2}
\end{eqnarray}
The fields $\left( \overline{\varphi }_{\mu }^{ac},\varphi _{\mu
}^{ac}\right) $ are a pair of complex conjugate bosonic fields. Each
field has $4\left( N^{2}-1\right) ^{2}$ components. Similarly, the
fields $\left( \overline{\omega }_{\mu }^{ac},\omega _{\mu
}^{ac}\right) $ are anticommuting. The local action (\ref{ll1}) is
renormalizable by power counting. More precisely, it has been shown
in \cite{Zwanziger:1989mf,Zwanziger:1992qr,Maggiore:1993wq} that the
Green functions obtained with the action $S_{0}$ with the insertion
of
the local composite operators $f^{abc}A_{\mu }^{a}\varphi _{\mu }^{bc}$ and $%
f^{abc}A_{\mu }^{a}\overline{\varphi }_{\mu }^{bc}$ are
renormalizable, the action $S_{0}$ being indeed renormalizable by a
multiplicative renormalization of the coupling constant $g$ and of
the fields \cite{Zwanziger:1989mf,Zwanziger:1992qr,Maggiore:1993wq}.
We remark that the action $S_{0}$ displays a global $U(f)$ symmetry,
$f=4\left( N^{2}-1\right) $, with respect to the composite index
$i=\left( \mu ,c\right) =1,...,f$,
of the additional fields $\left( \overline{%
\varphi }_{\mu }^{ac},\varphi _{\mu }^{ac},\overline{\omega }_{\mu
}^{ac},\omega _{\mu }^{ac}\right) $. Setting
\begin{equation}
\left( \overline{\varphi }_{\mu }^{ac},\varphi _{\mu }^{ac},\overline{\omega
}_{\mu }^{ac},\omega _{\mu }^{ac}\right) =\left( \overline{\varphi }%
_{i}^{a},\varphi _{i}^{a},\overline{\omega }_{i}^{a},\omega _{i}^{a}\right)
\;,  \label{l3}
\end{equation}
we get
\begin{eqnarray}
S_{0} &=&S_{\mathrm{YM}}+\int d^{4}x\;\left( b^{a}\partial_\mu A_\mu^{a}+\overline{c}%
^{a}\partial _{\mu }\left( D_{\mu }c\right) ^{a}\right) \;  \nonumber \\
&+&\int d^{4}x\left( \overline{\varphi }_{i}^{a}\partial _{\nu
}\left( D_{\nu }\varphi _{i}\right) ^{a}-\overline{\omega
}_{i}^{a}\partial _{\nu
}\left( D_{\nu }\omega _{i}\right) ^{a}\left. -g\left( \partial _{\nu }%
\overline{\omega }_{i}^{a}\right) f^{abm}\left( D_{\nu }c\right)
^{b}\varphi _{i}^{m}\right) \right) \;.    \label{l4}
\end{eqnarray}
For the $U(f)$ invariance we have
\begin{eqnarray}
U_{ij}S_{0}&=&0\;,\nonumber\\  \label{ll4} U_{ij}&=&\int
d^{4}x\left( \varphi _{i}^{a}\frac{\delta }{\delta \varphi
_{j}^{a}}-\overline{\varphi }_{j}^{a}\frac{\delta }{\delta
\overline{\varphi
}_{i}^{a}}+\omega _{i}^{a}\frac{\delta }{\delta \omega _{j}^{a}}-\overline{%
\omega }_{j}^{a}\frac{\delta }{\delta \overline{\omega
}_{i}^{a}}\right) \;.
\end{eqnarray}
The presence of the global $U(f)$ invariance means that one can make
use of the composite index $i=\left( \mu ,c\right) $. By means of
the diagonal operator $Q_{f}=U_{ii}$, the $i$-valued fields turn out
to possess an additional quantum number. As shown in
\cite{Zwanziger:1989mf,Zwanziger:1992qr,Maggiore:1993wq}, the action
$S_{0}$ is left invariant by the following nilpotent BRST\
transformations,
\begin{eqnarray}
sA_{\mu }^{a} &=&-\left( D_{\mu }c\right) ^{a}\;,  \nonumber \\
sc^{a} &=&\frac{1}{2}gf^{abc}c^{b}c^{c}\;,  \nonumber \\
s\overline{c}^{a} &=&b^{a}\;,\;\;\;\;\;\;sb^{a}=0\;,  \nonumber \\
s\varphi _{i}^{a} &=&\omega _{i}^{a}\;,\;\;\;\;\;s\omega _{i}^{a}=0\;,
\nonumber \\
s\overline{\omega }_{i}^{a} &=&\overline{\varphi }_{i}^{a}\;,\;\;\;\;\;s%
\overline{\varphi }_{i}^{a}=0\;,\;  \label{l6}
\end{eqnarray}
with
\begin{equation}
sS_{0}=0\;.  \label{l7}
\end{equation}
For further use, the quantum numbers of all fields entering the
action $S_{0}$ are displayed in Table 1.
\begin{table}[b]
  \centering
\begin{tabular}{|c|c|c|c|c|c|c|c|c|}
\hline & $A_{\mu }^{a}$ & $c^{a}$ & $\overline{c}^{a}$ & $b^{a}$ &
$\varphi
_{i}^{a} $ & $\overline{\varphi }_{i}^{a}$ & $\omega _{i}^{a}$ & $\overline{%
\omega }_{i}^{a}$ \\ \hline \hline \textrm{dimension} & $1$ & $0$ &
$2$ & $2$ & $1$ & $1$ & $1$ & $1$ \\ \hline
$\mathrm{ghost number}$ & $0$ & $1$ & $-1$ & $0$ & $0$ & $0$ & $1$ & $-1$ \\
\hline $Q_{f}\textrm{-charge}$ & $0$ & $0$ & $0$ & $0$ & $1$ & $-1$
& $1$ & $-1$
\\ \hline
\end{tabular}  \caption{Quantum numbers of the fields.}\label{tabel1}
\end{table}
It is worth noticing that, when $f^{abc}A_{\mu }^{a}\varphi _{\mu
}^{bc}$ and $f^{abc}A_{\mu }^{a}\overline{\varphi }_{\mu }^{bc}$
are treated as composite operators, they are introduced in the
starting action $S_{0}$ coupled to local external sources $M_{\mu
}^{ai}$,\ $V_{\mu }^{ai}$, namely
\begin{equation}
-\int d^{4}x\left( M_{\mu }^{ai}\left( D_{\mu }\varphi _{i}\right)
^{a}+V_{\mu }^{ai}\left( D_{\mu }\overline{\varphi }_{i}\right) ^{a}\right)
\;.  \label{l8}
\end{equation}
The horizon condition (\ref{m4}) is thus obtained from the quantum
action by requiring that, at the end of the computation, the sources
$M_{\mu }^{ai}$,\ $V_{\mu }^{ai}$ attain the physical values,
obtained by setting
\begin{equation}
M_{\mu \nu }^{ab}=V_{\mu \nu }^{ab}=\gamma ^{2}\delta ^{ab}\delta _{\mu \nu
}\;.  \label{l9}
\end{equation}
Indeed, expression (\ref{l8}) reduces precisely to
that of eq.(\ref{ll1}) when the sources $M_{\mu }^{ai}$,\ $%
V_{\mu }^{ai}$ attain their physical value.

\sect{Renormalization of the Zwanziger action in the presence of the
composite operator $A_{\mu }^{a}A_{\mu }^{a}$.} The purpose of this
section is to show that the renormalizability of the
local action $S_{0}\;$is preserved when, besides the operators $%
f^{abc}A_{\mu }^{a}\varphi _{\mu }^{bc}$ and $f^{abc}A_{\mu }^{a}\overline{%
\varphi }_{\mu }^{bc}$, also the local composite operator $A_{\mu
}^{a}A_{\mu }^{a}$ is introduced. This is a remarkable feature of the
Zwanziger action, allowing us to discuss the condensation of the operator $%
A_{\mu }^{a}A_{\mu }^{a}$ when the restriction to the Gribov region
$\Omega $ is implemented. To discuss the renormalizability of the
model in the presence of $A_{\mu }^{2}$, we start from the following
complete action
\begin{equation}
\Sigma =S_{0}+S_{\mathrm{s}}+S_{\mathrm{ext}}\;,  \label{r1}
\end{equation}
where $S_{\mathrm{s}}$ is the term containing all needed local
composite operators with their respective local sources, and is
given by
\begin{equation}
S_{\mathrm{s}}=s\int d^{4}x\left( -U_{\mu }^{ai}\left( D_{\mu
}\varphi _{i}\right) ^{a}-V_{\mu }^{ai}\left( D_{\mu
}\overline{\omega }_{i}\right) ^{a}-U_{\mu
}^{ai}V_{\mu }^{ai}+\frac{1}{2}\eta A_{\mu }^{a}A_{\mu }^{a}-\frac{1}{2}%
\zeta \tau \eta \right) \;,  \label{r2}
\end{equation}
where the BRST operator acts as
\begin{eqnarray}
sU_{\mu }^{ai} &=&M_{\mu }^{ai}\;,\;\;\;\;sM_{\mu }^{ai}=0\;,  \nonumber \\
sV_{\mu }^{ai} &=&N_{\mu }^{ai}\;,\;\;\;\;sN_{\mu }^{ai}=0\;,  \label{r3}
\end{eqnarray}
and
\begin{equation}
s\eta =\tau \;,\;\;\;\;s\tau =0\;.  \label{r4}
\end{equation}
\begin{table}
  \centering
  \begin{tabular}{|c|c|c|c|c|c|c|c|c|}
\hline
& $U_{\mu }^{ai}$ & $M_{\mu }^{ai}$ & $N_{\mu }^{ai}$ & $V_{\mu }^{ai}$ & $%
\eta $ & $\tau $ & $K_{\mu }^{a}$ & $L^{a}$
\\ \hline\hline \textrm{dimension} & $2$ & $2$ & $2$ &
$2$ & $2$ & $2$ & $3$ & $4$ \\ \hline $\mathrm{ghost number}$ & $-1$
& $0$ & $1$ & $0$ & $-1$ & $0$ & $-1$ & $-2$ \\ \hline
$Q_{f}\textrm{-charge}$ & $-1$ & $-1$ & $1$ & $1$ & $0$ & $0$ & $0$ & $0$ \\
\hline
\end{tabular}
\caption{Quantum numbers of the sources.}\label{tabel2}
\end{table}
Therefore, for $S_{\mathrm{s}}$ one gets
\begin{eqnarray}
S_{\mathrm{s}} &=&\int d^{4}x\left( -M_{\mu }^{ai}\left( D_{\mu
}\varphi _{i}\right) ^{a}-gU_{\mu }^{ai}f^{abc}\left( D_{\mu
}c\right) ^{b}\varphi _{i}^{c}+U_{\mu }^{ai}\left( D_{\mu }\omega
_{i}\right) ^{b}\right.
\nonumber \\
&-&N_{\mu }^{ai}\left( D_{\mu }\overline{\omega }_{i}\right)
^{a}-V_{\mu }^{ai}\left( D_{\mu }\overline{\varphi }_{i}\right)
^{a}+gV_{\mu }^{ai}f^{abc}\left( D_{\mu }c\right)
^{b}\overline{\omega }_{i}^{c}
\nonumber \\
 &-&\left.M_{\mu }^{ai}V_{\mu }^{ai}+U_{\mu }^{ai}N_{\mu }^{ai}+\frac{1}{2}%
\tau A_{\mu }^{a}A_{\mu }^{a}+\eta A_{\mu }^{a}\partial _{\mu }c^{a}-\frac{1%
}{2}\zeta \tau ^{2}\right) \;.  \label{r5}
\end{eqnarray}
As already noticed, the sources $M_{\mu }^{ai},$ $V_{\mu }^{ai}$ are needed
to introduce the composite operators $\left( D_{\mu }\varphi _{i}\right)
^{a} $ and $\left( D_{\mu }\overline{\varphi }_{i}\right) ^{a}$. The sources
$U_{\mu }^{ai}$, $N_{\mu }^{ai}$ define the BRST variations of these
operators, given by $\left( D_{\mu }\omega _{i}\right) ^{b}$ and $\left( D_{\mu }%
\overline{\omega }_{i}\right) ^{a}$. The physical value of these sources is
given by
\begin{eqnarray}
M_{\mu \nu }^{ab} &=&V_{\mu \nu }^{ab}=\gamma ^{2}\delta ^{ab}\delta _{\mu
\nu }\;,  \nonumber \\
U_{\mu \nu }^{ab} &=&N_{\mu \nu }^{ab}=0\;.  \label{r6}
\end{eqnarray}
The local composite operator $A_{\mu }^{a}A_{\mu }^{a}$ and its BRST
variation, $A_{\mu }^{a}\partial _{\mu }c^{a}$, are then introduced by means
of the local sources $\tau $, $\eta $. We also notice that the complete
action $\Sigma $ contains terms quadratic in the external sources, namely $%
\left( M_{\mu }^{ai}V_{\mu }^{ai}-U_{\mu }^{ai}N_{\mu }^{ai}\right) $ and $%
\zeta \tau ^{2}$. These terms, allowed by power counting, are in fact needed
for the multiplicative renormalizability of the model. As shown in \cite{Verschelde:2001ia}%
, the dimensionless LCO parameter $\zeta $ of the quadratic term in the source $%
\tau $ is needed to account for the divergences present in the
correlation function $\left\langle
A_\mu^{2}(x)A_\nu^{2}(y)\right\rangle $ for $x\rightarrow y$. It
should be remarked that, unlike for the term quadratic in the
external source $\tau$, we have not introduced a new free parameter
for the quadratic term $\left( M_{\mu }^{ai}V_{\mu }^{ai}-U_{\mu
}^{ai}N_{\mu }^{ai}\right)$ in expression (\ref{r5}). As we shall
see, this term goes through the renormalization without the need of
introducing a new parameter for its renormalizability. This is a
remarkable feature of the Zwanziger action which plays an important
role when the ghost propagator in the presence of the Gribov horizon
will be discussed, see section 6.\\\\Finally, the term
$S_{\mathrm{ext}}$ is the source term needed to define the nonlinear
BRST transformations of the gauge and ghost fields, i.e.
\begin{equation}
S_{\mathrm{ext}}=\int d^{4}x\left( -K_{\mu }^{a}\left( D_{\mu }c\right) ^{a}+\frac{1}{%
2}gL^{a}f^{abc}c^{b}c^{c}\right) \;.  \label{r7}
\end{equation}
The technical details concerning the algebraic renormalization
procedure have been worked out in Appendix A. In summary, the
Zwanziger action in the presence of the local operator
$A_{\mu}^{a}A_{\mu }^{a}$ is multiplicative renormalizable. In turn,
this ensures that the quantum effective action obeys the homogeneous
renormalization group equations (RGE). This is an important feature
of the model, which will be useful when we shall try to obtain
estimates for both the Gribov and mass parameter.\\\\The effective
action is defined upon setting the sources $U_{\mu \nu }^{ab}$,
$N_{\mu \nu }^{ab}$, $K_{\mu }^{a}$, $L^{a}$ and $\eta$ equal to
zero and implementing the condition (\ref{l9}). Doing so, we get
\begin{eqnarray}
S &=&S_{0}+S_\gamma+\int d^{4}x\left( \frac{\tau }{2}A_{\mu
}^{a}A_{\mu
}^{a}-\frac{\zeta }{2}\tau ^{2}\right) \;, \nonumber \label{f1} \\
S_\gamma  &=&\int d^{4}x\left[ -\gamma ^{2}gf^{abc}A_{\mu
}^{a}\varphi _{\mu }^{bc}-\gamma ^{2}gf^{abc}A_{\mu
}^{a}\overline{\varphi }_{\mu }^{bc}-4\left( N^{2}-1\right) \gamma
^{4}\right]\;.
\end{eqnarray}
The term $-4\left( N^{2}-1\right) \gamma ^{4}$ originates from the
quadratic term in the external sources, namely $\left( -M_{\mu
}^{ai}V_{\mu }^{ai}+U_{\mu }^{ai}N_{\mu }^{ai}\right) $, in
expression (\ref{r5}), evaluated at the physical values given by
eq.(\ref{l9}).
\\\\Following \cite{Verschelde:2001ia,Browne:2003uv,Knecht:2001cc,Dudal:2003by}, we
introduce a Hubbard-Stratonovich field $\sigma$ by means of the
following unity
\begin{equation}
 \label{unity}1=\int[d\sigma]e^{-\frac{1}{2\zeta}\int d^4x\left[\frac{\sigma}{g}+\frac{1}{2}A_\mu^a
 A_\mu^a-\zeta\tau\right]^2}\;,
\end{equation}
to remove the term proportional to $\tau^2$. The source $\tau$ is
henceforth linearly coupled to the field $\sigma$, as can be
directly seen from the action, which now reads
\begin{eqnarray}
S &=&S_{0}+S_\gamma +S_\sigma+\int d^{4}x\left( -\tau \frac{\sigma }{g}\right) \;,  \label{f1bis} \nonumber\\
S_\sigma&=&\frac{\sigma
^{2}}{2g^{2}\zeta }+\frac{1}{2}\frac{g\sigma }{g^{2}\zeta }A_{\mu}^aA_{\mu}^a+%
\frac{1}{8\zeta }\left( A_{\mu}^aA_\mu^a\right) ^{2}\;.
\end{eqnarray}
The following identification is easily derived
\cite{Verschelde:2001ia,Browne:2003uv,Knecht:2001cc,Dudal:2003by}
\begin{equation}\label{new1}
\left\langle A_\mu^a A_\mu^
a\right\rangle=-\frac{1}{g}\left\langle\sigma\right\rangle\;,
\end{equation}
from which it follows that a nonvanishing vacuum expectation value
of the field $\sigma$ will result in a nonvanishing condensate
$\left\langle A_\mu^a A_\mu^ a\right\rangle$. \\\\The quantum action
$\Gamma$ is obtained through the definition
\begin{equation}
e^{-\Gamma}=\int \left[d\Phi\right]e^{-S_{0}-S_\gamma -S_\sigma }\;,
\label{3}
\end{equation}
where $\Phi$ is a shorthanded notation for all the relevant
fields.\\\\The value for $\left\langle \sigma \right\rangle $ is
found through the minimization condition
\begin{equation}
\frac{\p\Gamma}{\p\sigma }=0\;.  \label{5}
\end{equation}
The horizon is implemented by the condition
\cite{Zwanziger:1989mf,Zwanziger:1992qr}.
\begin{equation}
\frac{\p \Gamma }{\p\gamma^2 }=0\;.  \label{new2}
\end{equation}
Let us show this here. The following equivalence is readily found
\begin{equation}\label{invoegsel1}
    \frac{\p\Gamma}{\p\gamma^2}=0\Leftrightarrow\left\langle
    gf^{abc}A_\mu^a\varphi_{\mu}^{bc}\right\rangle+\left\langle
    gf^{abc}A_\mu^a\overline{\varphi}_{\mu}^{bc}\right\rangle=-8\left(N^2-1\right)\gamma^2\;,
\end{equation}
>From expressions (\ref{m1}) and (\ref{ll1}), it follows that
\begin{equation}\label{invoegsel3}
    -2\gamma^2\left\langle h\right\rangle=\left\langle
    gf^{abc}A_\mu^a\varphi_{\mu}^{bc}\right\rangle+\left\langle
    gf^{abc}A_\mu^a\overline{\varphi}_{\mu}^{bc}\right\rangle\;.
\end{equation}
The combination of eq.(\ref{invoegsel1}) with eq.(\ref{invoegsel3})
gives rise to the horizon condition eq.(\ref{m4}). In order to
conclude this, it is tacitly assumed that $\gamma\neq0$. We notice
that the condition (\ref{new2}) does possess the solution
$\gamma=0$. This is an artefact of the reformulation of the horizon
condition in terms of the equation (\ref{new2}), and must be
excluded as it does not lead to the horizon condition (\ref{m4}). We
shall, however, continue to keep this solution of the gap equation
(\ref{new2}), as $\gamma\equiv0$ corresponds to the case where the
restriction to the Gribov region $\Omega$ would not be implemented.
In this case, we must only solve the gap equation stemming from
eq.(\ref{5}) with $\gamma\equiv0$.\\\\The original Gribov-Zwanziger
model, i.e. without the inclusion of the operator $A_\mu^2$, is
obtained by only retaining the condition (\ref{new2}) with
$\sigma\equiv0$.\\\\Up to now, the LCO parameter $\zeta$ is still a
free parameter of the theory. We do not intend here to give a
complete overview of the LCO formalism, we suffice by saying that
$\zeta$ is fixed by the demand that the action $\Gamma$ should obey
the homogeneous renormalization group equation
\begin{equation}\label{LCOrge}
    \left(\omu\frac{\p}{\p\omu}+\beta(g^2)\frac{\p}{\p g^2}
    +\gamma_{\gamma^2}(g^2)\gamma^2\frac{\p}{\p
    \gamma^2}+\gamma_\sigma(g^2)\sigma\frac{\p}{\p\sigma}\right)\Gamma=0\;,
\end{equation}
with
\begin{eqnarray}\label{LCOrge2}
\omu\frac{\p g^2}{\p\omu}&=&\beta(g^2)\;,\nonumber\\
  \omu\frac{\p\gamma^2}{\p \omu} &=& \gamma_{\gamma^2}(g^2)\gamma^2 \;,\nonumber\\
   \omu\frac{\p \sigma}{\p \omu} &=&
   \gamma_{\sigma}(g^2)\sigma\;.
\end{eqnarray}
This can be accomodated for by making $\zeta$ a function of the
running coupling constant $g^2$, in which case it is found that
\begin{equation}\label{LCOrge3}
    \zeta(g^2)=\frac{\zeta_0}{g^2}+\zeta_1+\zeta_2 g^2+\cdots\;.
\end{equation}
We refer to the available literature
\cite{Verschelde:2001ia,Browne:2003uv,Knecht:2001cc,Dudal:2003gu,Dudal:2003by,Dudal:2004rx}
for a detailed account of the LCO formalism.

\subsection{Renormalization group invariance of the one-loop effective
action in the $\MSbar$ scheme without the inclusion of $A_\mu^2$.}
Before proceeding with the detailed analysis of the horizon
condition in the presence of the local operator $A_{\mu }^{a}A_{\mu
}^{a}$, let us first derive the horizon condition and check the
explicit renormalization group invariance of the quantum action
$\Gamma$ by switching off the source $\tau $ coupled to the operator
$A_{\mu }^{a}A_{\mu }^{a}$. This amounts to consider the original
Gribov-Zwanziger model. We consider thus the action
\begin{eqnarray}
S &=&S_{0}+S_\gamma\;. \label{new3}
\end{eqnarray}
The one-loop effective action $\Gamma^{(1)}$ is easily obtained from
the quadratic part of eq.(\ref{new3})
\begin{equation}
e^{-\Gamma_\gamma ^{(1)} }=\int \left[ D\Phi \right] e^{-S_{\mathrm{quad}%
}}\;,  \label{co10}
\end{equation}
with $S_{\mathrm{quad}}$ given by
\begin{eqnarray}
S_{\mathrm{quad}} &=&\int d^{4}x\left[ \frac{1}{4}\left( \partial _{\mu
}A_{\nu }^{a}-\partial _{\nu }A_{\mu }^{a}\right) ^{2}+\frac{1}{2\alpha }%
\left( \partial_\mu A_\mu^{a}\right) ^{2}+\overline{\varphi }_{\mu
}^{ab}\partial
^{2}\varphi _{\mu }^{ab}\right.  \nonumber \\
&-&\left. \gamma ^{2}g\left( f^{abc}A_{\mu }^{a}\varphi _{\mu
}^{bc}+f^{abc}A_{\mu }^{a}\overline{\varphi }_{\mu }^{bc}\right)
-4(N^{2}-1)\gamma ^{4}\vphantom{\frac{1}{4}\left( \partial _{\mu
}A_{\nu }^{a}-\partial _{\nu }A_{\mu }^{a}\right) ^{2}}\right] \;,
\label{co11}
\end{eqnarray}
where the limit $\alpha \rightarrow 0$ is understood in order to
recover the Landau gauge.  After a straightforward computation, one
gets
\begin{equation}
\Gamma^{(1)} =-4(N^{2}-1)\gamma ^{4}+\frac{(N^{2}-1)}{2}\left(
d-1\right) \int \frac{d^{d}p}{\left( 2\pi \right) ^{d}}\ln \left(
p^{4}+2Ng^{2}\gamma ^{4}\right) \;. \label{co12}
\end{equation}
Dimensional regularization, with $d=4-\varepsilon$, will be employed
throughout this work. Taking the derivative of $\Gamma^{(1)} $, one
reobtains the original gap equation for the Gribov parameter $\gamma
$, namely
\begin{equation}
\frac{\p \Gamma^{(1)} }{\p\gamma }=0 \Rightarrow 1=%
\frac{N\left( d-1\right) }{4}g^{2}\int \frac{d^{d}p}{\left( 2\pi \right) ^{d}%
}\frac{1}{\left( p^{4}+2Ng^{2}\gamma ^{4}\right) }\;.  \label{co13}
\end{equation}
More precisely, recalling that
\begin{equation}
\int \frac{d^{d}p}{\left( 2\pi \right) ^{d}}\ln \left( p^{4}+\rho
^{2}\right) =-\frac{\rho ^{2}}{32\pi ^{2}}\left( \ln \frac{\rho ^{2}}{%
\overline{\mu }^{4}}-3\right) +\frac{1}{\varepsilon }\frac{4\rho
^{2}}{32\pi ^{2}}\;,  \label{co14}
\end{equation}
the one-loop effective action $\Gamma^{(1)}$ reads
\begin{equation}
\Gamma^{(1)}=-4(N^{2}-1)\gamma ^{4}-\frac{3(N^{2}-1)}{64\pi ^{2}}%
\left( 2Ng^{2}\gamma ^{4}\right) \left( \ln \frac{2Ng^{2}\gamma ^{4}}{%
\overline{\mu }^{4}}-\frac{5}{3}\right) \;,  \label{co15}
\end{equation}
where the $\MSbar$ renormalization scheme has been used. \\\\In
order to check the renormalization group invariance of
$\Gamma^{(1)}$, we need to know the anomalous dimension of the
Gribov parameter $\gamma $. This is easily obtained from
eq.(\ref{co6}), yielding
\begin{equation}
\gamma_{\gamma ^{2}}(g^2)=-\frac{1}{2}\left(
\frac{\beta(g^2)}{2g^2}-\gamma _{A}(g^2)\right) \;,  \label{co16}
\end{equation}
where $\gamma _{A}(g^2)$ stands for the anomalous dimension of the gauge field $%
A_{\mu }^{a}$. Thus, at one-loop order,
\begin{equation}
\overline{\mu }\frac{d\Gamma ^{(1)} }{d\overline{\mu }}=\left(
4(N^{2}-1)\left( \frac{\beta^{(1)}(g^2)}{2g^2}-\gamma _{A}^{(1)}(g^2)\right) +\frac{%
3(N^{2}-1)}{16\pi ^{2}}2Ng^{2}\right) \gamma ^{4}\;.  \label{co17}
\end{equation}
Furthermore, from (see e.g. \cite{Gracey:2002yt})
\begin{eqnarray}
\beta^{(1)}(g^2) &=&-\frac{22}{3}\frac{g^{4}N}{16\pi ^{2}}\;,
\nonumber \\
\gamma _{A}^{(1)}(g^2) &=&-\frac{13}{6}\frac{g^{2}N}{16\pi
^{2}}\;, \label{co18}
\end{eqnarray}
it follows
\begin{equation}
\overline{\mu }\frac{d\Gamma ^{(1)}}{d\overline{\mu }}=0\;,
\label{co19}
\end{equation}
which establishes the RGE invariance of the effective action at
the order considered. \\\\We are now
ready to face the more complex case in which the local composite operator $%
A_{\mu }^{a}A_{\mu }^{a}$ is present. This will be the topic of the next
section.

\sect{One-loop effective action in the $\MSbar$ scheme with the
inclusion of $A_\mu^2$.}
\subsection{Calculation of the one-loop effective potential.}
Let us turn to the explicit one-loop evaluation of the effective
action $\Gamma$ in the presence of $A_\mu^2$. At one-loop, it turns
out that\footnote{We shall drop from now on the superscript $^{(1)}$
indicating that we are working at one-loop order.}
\begin{equation}
\Gamma=-4\left( N^{2}-1\right) \gamma ^{4}+\frac{\sigma ^{2}}{2g^{2}\zeta }+%
\frac{N^{2}-1}{2}\ln \det \left[ p^{2}\delta _{\mu \nu }+\frac{2Ng^{2}\gamma
^{4}}{p^{2}}\delta _{\mu \nu }-p_{\mu }p_{\nu }\left( 1-\frac{1}{\alpha }%
\right) +\frac{g\sigma }{g^{2}\zeta }\delta _{\mu \nu }\right]\;,
\label{6}
\end{equation}
or
\begin{equation}
\Gamma =-4\left( N^{2}-1\right) \gamma ^{4}+\frac{\sigma ^{2}}{2g^{2}\zeta }+%
\frac{N^{2}-1}{2}(d-1)\int \frac{d^{d}p}{\left( 2\pi \right)
^{d}}\ln \left[ p^{4}+2Ng^{2}\gamma ^{4}+\frac{g\sigma }{g^{2}\zeta
}p^{2}\right]\;.   \label{7}
\end{equation}
Before calculating the integral, we quote the two gap equations
\begin{eqnarray}
\frac{\p\Gamma}{\p\sigma }=0 &\Leftrightarrow &\frac{\sigma }{\zeta _{0}}%
\left( 1-\frac{\zeta _{1}}{\zeta _{0}}g^{2}\right) +\frac{\left(
N^{2}-1\right) }{2}\frac{g(d-1)}{\zeta _{0}}\int
\frac{d^{d}p}{\left( 2\pi
\right) ^{d}}\frac{p^{2}}{p^{4}+\frac{g\sigma }{\zeta _{0}}%
p^{2}+2Ng^{2}\gamma^{4}}=0 \;, \nonumber  \label{7bis} \\
\frac{\p \Gamma }{\p \gamma }=0 &\Leftrightarrow &\gamma^3=\gamma^3\frac{d-1}{4}%
g^{2}N\int \frac{d^{d}p}{\left( 2\pi \right) ^{d}}\frac{1}{p^{4}+\frac{%
g\sigma }{\zeta _{0}}p^{2}+2Ng^{2}\gamma^{4}}\;.
\end{eqnarray}
The second gap equation of (\ref{7bis}), being the horizon
condition, gives rise to the one obtained in the previous paper
\cite{Sobreiro:2004us}, while the first one describes the
condensation of $A_\mu^2$ when the restriction to the Gribov region
$\Omega$ is implemented. We notice that that the explicit value of
the Gribov parameter $\gamma$ is influenced by the presence of
$\left\langle A_\mu^2\right\rangle$.
\\\\
It remains to calculate
\begin{equation}
\mathcal{I}=\int \frac{d^{d}p}{\left( 2\pi \right) ^{d}}\ln \left[
p^{4}+bp^{2}+c\right]\;,   \label{8}
\end{equation}
with
\begin{equation}
b =\frac{g\sigma}{\zeta_0}\;,\;\;\;\;\;\;\;\; c=2Ng^2\gamma^4\;,
\label{8bis}
\end{equation}
Since
\begin{equation}
p^{4}+bp^{2}+c =\left( p^{2}+\omega _{1}\right) \left( p^{2}+\omega
_{2}\right)   \;,  \label{9}
\end{equation}
with
\begin{equation}
\omega _{1} =\frac{b+\sqrt{b^{2}-4c}}{2}\;,\;\;\;\;\;\;\;\; \omega
_{2} =\frac{b-\sqrt{b^{2}-4c}}{2}\;, \label{9b}
\end{equation}
one has
\begin{equation}
\mathcal{I}=\int \frac{d^{d}p}{\left( 2\pi \right) ^{d}}\ln \left(
p^{2}+\omega _{1}\right) +\int \frac{d^{d}p}{\left( 2\pi \right)
^{d}}\ln \left( p^{2}+\omega _{2}\right)\;.   \label{10}
\end{equation}
To make sense, the expression (\ref{8}) should be real to ensure
that the one-loop effective action is real-valued. Therefore, we
must demand that $c\geq0$. If $b\geq0$, $\mathcal{I}$ is certainly
real. However, when $b^2-4c\leq0$, then also $b<0$ is allowed. We
should thus have a positive Gribov parameter $\gamma^4$, while the
condensate $\left\langle A_\mu^2\right\rangle$ can be negative or
positive, depending on the case.\\\\Using
\begin{equation}
\int \frac{d^{d}p}{\left( 2\pi \right) ^{d}}\ln \left( p^{2}+m^{2}\right) =%
\frac{-m^{4}}{32\pi ^{2}}\left( \frac{2}{\varepsilon }-\ln \frac{m^{2}}{%
\overline{\mu }^{2}}+\frac{3}{2}\right) \;,  \label{11}
\end{equation}
it holds
\begin{equation}
\mathcal{I}=-\frac{\omega _{1}^{2}}{32\pi ^{2}}\left( \frac{2}{\varepsilon }-\ln \frac{%
\omega _{1}}{\overline{\mu }^{2}}+\frac{3}{2}\right) -\frac{\omega _{2}^{2}%
}{32\pi ^{2}}\left( \frac{2}{\varepsilon }-\ln \frac{\omega _{2}}{\overline{%
\mu }^{2}}+\frac{3}{2}\right)\;.   \label{12}
\end{equation}
Finally, in the $\MSbar$ scheme, we obtain
\begin{eqnarray}\label{new4}
\Gamma&=&-4\left( N^{2}-1\right) \gamma ^{4}+\frac{\sigma ^{2}}{2\zeta _{0}%
}\left( 1-\frac{\zeta _{1}}{\zeta _{0}}g^{2}\right) +\frac{3\left(
N^{2}-1\right) }{2}\times   \nonumber  \label{14} \\
&&\left[ \frac{\left( \frac{g\sigma }{\zeta _{0}}+\sqrt{\frac{g^{2}\sigma
^{2}}{\zeta _{0}^{2}}-8g^{2}N\gamma ^{4}}\right) ^{2}}{128\pi ^{2}}\left(
\ln \frac{\frac{g\sigma }{\zeta _{0}}+\sqrt{\frac{g^{2}\sigma ^{2}}{\zeta
_{0}^{2}}-8g^{2}N\gamma ^{4}}}{2\overline{\mu }^{2}}-\frac{5}{6}\right)
\right.   \nonumber \\
&+&\left. \frac{\left( \frac{g\sigma }{\zeta
_{0}}-\sqrt{\frac{g^{2}\sigma ^{2}}{\zeta _{0}^{2}}-8g^{2}N\gamma
^{4}}\right) ^{2}}{128\pi ^{2}}\left( \ln \frac{\frac{g\sigma
}{\zeta _{0}}-\sqrt{\frac{g^{2}\sigma ^{2}}{\zeta
_{0}^{2}}-8g^{2}N\gamma ^{4}}}{2\overline{\mu
}^{2}}-\frac{5}{6}\right) \right]\;.
\end{eqnarray}
To lighten the notation a bit, let us introduce the new
variables\footnote{In comparison with the previous article
\cite{Sobreiro:2004us}, we have the correspondence
$\lambda^4=4\gamma^4$ with the Gribov parameter $\gamma^4$ as
defined there. It is actually this $\gamma^4$ which will enter the
modified propagators, see \cite{Sobreiro:2004us} and further in
this paper.}
\begin{eqnarray}\label{newvar}
\lambda^4&=&8g^2N\gamma^4\;,\\
m^2&=&\frac{g\sigma}{\zeta_0}\;.\label{def1}
\end{eqnarray}
in which case the action (\ref{new4}) can be rewritten as
\begin{eqnarray}\label{new5}
\Gamma  &=&- \frac{\left( N^{2}-1\right)\lambda
^{4}}{2g^2N}+\frac{\zeta_0m^4}{2g^2}\left( 1-\frac{\zeta
_{1}}{\zeta _{0}}g^{2}\right) \nonumber\\&+&\frac{3\left(
N^{2}-1\right)
}{256\pi^2}\left[\left(m^2+\sqrt{m^4-\lambda^4}\right)^2
\left(\ln\frac{m^2+\sqrt{m^4-\lambda^4}}{2\omu^2}-\frac{5}{6}\right)\right.\nonumber\\
&+&\left.\left(m^2-\sqrt{m^4-\lambda^4}\right)^2
\left(\ln\frac{m^2-\sqrt{m^4-\lambda^4}}{2\omu^2}-\frac{5}{6}\right)\right]\;.
\end{eqnarray}
We notice that the foregoing expression is also valid, i.e.
real-valued, in the case in which $m^4<\lambda^4$, as
$\ell_+(m,\lambda)$ and $\ell_-(m,\lambda)$, defined by,
\begin{eqnarray}\label{nieuw1}
    \ell_+(m,\lambda)&=&\left(m^2+\sqrt{m^4-\lambda^4}\right)^2
\left(\ln\frac{m^2+\sqrt{m^4-\lambda^4}}{2\omu^2}-\frac{5}{6}\right)\nonumber\\
\ell_-(m,\lambda)&=&\left(m^2-\sqrt{m^4-\lambda^4}\right)^2
\left(\ln\frac{m^2-\sqrt{m^4-\lambda^4}}{2\omu^2}-\frac{5}{6}\right)
\end{eqnarray}
are complex conjugate\footnote{Using $\ln(z) = \ln|z|+i\arg(z)$
with $-\pi<\arg(z)\leq\pi$.}.
\\\\The horizon condition, eq.(\ref{new2}), can be translated to
\begin{equation}\label{new6}
    \frac{\p\Gamma}{\p\lambda}=0\;,
\end{equation}
and the gap equation (\ref{5}) to
\begin{equation}
\frac{\p\Gamma}{\p m^2 }=0  \label{new7}\;.
\end{equation}
As a check of this one-loop calculation, the expression (\ref{new5})
with $m^2\equiv0$ reduces to the result obtained earlier in
eq.(\ref{co15}), i.e. the original Gribov-Zwanziger model without
the inclusion of $A_\mu^2$. If $\lambda\equiv0$, i.e. the case where
the condensation of $A_\mu^2$ is investigated without implementing
the restriction to the Gribov region $\Omega$, eq.(\ref{new5})
coincides with the result of
\cite{Verschelde:2001ia,Browne:2003uv,Dudal:2003by}. From
\cite{Dudal:2002pq}, one knows that
\begin{eqnarray}\label{new19bis}
\omu\frac{\p \left\langle
A_\mu^2\right\rangle}{\p\omu}&=&\gamma_{A_\mu^2}(g^2) \left\langle
A_\mu^2\right\rangle
=-\left(\frac{\beta(g^2)}{2g^2}+\gamma_A(g^2)\right) \left\langle
A_\mu^2\right\rangle\;,
\end{eqnarray}
or, using the relation (\ref{new1}) and the definition (\ref{def1}),
\begin{eqnarray}\label{new19}
\omu\frac{\p
m^2}{\p\omu}&=&\gamma_{m^2}(g^2)m^2=\left(\frac{\beta(g^2)}{2g^2}-\gamma_A(g^2)\right)m^2\;,
\end{eqnarray}
while from eq.(\ref{co16}), it can be inferred that
\begin{equation}\label{new20}
\overline{\mu }\frac{\p\lambda}{\p\overline{\mu
}}=\gamma_\lambda(g^2)\lambda=\frac{1}{4}\left(
\frac{\beta(g^2)}{2g^2}+\gamma _{A}(g^2)\right) \lambda\;.
\end{equation}
We notice the remarkable fact that the anomalous dimensions of the
Gribov parameter and of the operator $A_\mu^2$ are proportional to
each other, to all orders of perturbation theory.\\\\It can now be
checked that $\Gamma$ is renormalization group invariant, namely
\begin{equation}\label{new28}
    \omu\frac{d}{d\omu}\Gamma=0\;.
\end{equation}
Finally, taking the derivatives of the action given in
eq.(\ref{new5}) gives rise to
\begin{eqnarray}
\label{new8}
\frac{1}{\lambda^3}\frac{\p\Gamma}{\p\lambda}&=&-\frac{2\left(N^2-1\right)}{g^2N}+\frac{3\left(N^2-1\right)}{256\pi^2}
\left[-4\frac{\left(m^2+\sqrt{m^4-\lambda^4}\right)}{\sqrt{m^4-\lambda^4}}\ln\frac{m^2+\sqrt{m^4-\lambda^4}}{2\omu^2}\nonumber\right.\\
&+&\left.4\frac{\left(m^2-\sqrt{m^4-\lambda^4}\right)}{\sqrt{m^4-\lambda^4}}\ln\frac{m^2-\sqrt{m^4-\lambda^4}}{2\omu^2}+\frac{8}{3}\right]\;,
\end{eqnarray}
and
\begin{eqnarray}
\label{new9} \frac{\p\Gamma}{\p m^2}&=&\nonumber\\
\frac{\zeta_0m^2}{g^2}\left(1-\frac{\zeta_1}{\zeta_0}g^2\right)&+&\frac{3\left(N^2-1\right)}{256\pi^2}
\left[2\left(m^2+\sqrt{m^4-\lambda^4}\right)\left(1+\frac{m^2}{\sqrt{m^4-\lambda^4}}\right)\ln\frac{m^2+\sqrt{m^4-\lambda^4}}{2\omu^2}\right.\nonumber\\
&+&\left.2\left(m^2-\sqrt{m^4-\lambda^4}\right)\left(1-\frac{m^2}{\sqrt{m^4-\lambda^4}}\right)\ln\frac{m^2-\sqrt{m^4-\lambda^4}}{2\omu^2}-\frac{8}{3}m^2\right]\;.\nonumber\\
\end{eqnarray}
\subsection{Solving the gap equations.}
We have now all the ingredients at hand to search for estimates of
the mass parameter $m^2$ and Gribov parameter $\lambda$ as solutions
of the gap equations (\ref{new8}) and (\ref{new9}). To avoid
misinterpretations due to the suggestive use of the notation $m^2$,
we remark that, due to the presence of $\lambda$, the mass parameter
does not even appear as a pole in the tree level gluon propagator,
see eq.(\ref{prop2}). \\\\Let us first consider the pure
Gribov-Zwanziger case, i.e. we put $m^2\equiv0$ in the expression
(\ref{new5}). The relevant gap equation (horizon condition) reads
\begin{equation}\label{new10}
    \frac{\p\Gamma}{\p\lambda}=\lambda^3\left(-\frac{2\left(N^2-1\right)}{g^2N}-\frac{3\left(N^2-1\right)}{64\pi^2}\left(\ln\frac{\lambda^4}{4\omu^4}-\frac{5}{3}\right)-\frac{3\left(N^2-1\right)}{64\pi^2}\right)=0\;.
\end{equation}
We remind here that the solution $\lambda=0$ must be rejected. The
natural choice for the renormalization scale is to set
$\omu^2=\frac{\lambda^4}{4}$ to kill the logarithms, and we find
\begin{equation}\label{new11}
\left.\frac{g^2N}{16\pi^2}\right|_{\omu^2=\frac{\lambda^2}{2}}=4\;.
\end{equation}
In principle, from
\begin{equation}\label{new12}
    g^2(\omu^2)=\frac{1}{\beta_0\ln\frac{\omu^2}{\lms^2}}\;,\;\;\;\;\;\textrm{with
    }\;\;\;\;\beta_0=\frac{11}{3}\frac{N}{16\pi^2}\;,
\end{equation}
eq.(\ref{new11}) could be used to determine an estimate for the
Gribov parameter,  however it might be clear that this is
meaningless since the corresponding expansion parameter
(\ref{new11}) is far too big.\\\\It is interesting to notice that,
in a general massless renormalization scheme, the one-loop action
with $m^2\equiv0$ would read
\begin{equation}\label{extraatje1}
    \Gamma=-\frac{\left(N^2-1\right)}{2g^2N}\lambda^4-\frac{3\lambda^4\left(N^2-1\right)}{264\pi^2}\left(\ln\frac{\lambda^4}{4\omu^4}+a\right)\;,
\end{equation}
with $a$ an arbitrary constant. The corresponding gap equation
equals
\begin{equation}\label{extraatje2}
    \frac{\p\Gamma}{\p\lambda}=\lambda^3\left(-\frac{2\left(N^2-1\right)}{g^2N}-\frac{3\left(N^2-1\right)}{64\pi^2}\left(\ln\frac{\lambda^4}{4\omu^4}+a\right)-\frac{3\left(N^2-1\right)}{64\pi^2}\right)=0\;.
\end{equation}
Denoting by $\lambda_*$ a solution of eq.(\ref{extraatje2}), for the
vacuum energy corresponding to (\ref{extraatje1}) one finds
\begin{equation}\label{new12bis}
    E_\mathrm{vac}= \Gamma(\lambda_*)=\frac{3(N^2-1)}{64\pi^2}\frac{\lambda_*^4}{4}\;.
\end{equation}
This expression is valid for all $\omu$ and for all $a$. The vacuum
energy is thus always nonnegative at one-loop order in the original
Gribov-Zwanziger model.
\\\\The gap equation (\ref{new9}) with $\lambda\equiv0$ obviously
has the solution already obtained in
\cite{Verschelde:2001ia,Browne:2003uv,Dudal:2003by} where the
restriction to the Gribov region $\Omega$ was not taken into
account. We recall the values
\begin{eqnarray}
\label{rge32a}\frac{g^2N}{16\pi^2}&=&\frac{36}{187}\approx0.193\; ,\\
\label{rge32b}m^2&=&e^{\frac{17}{12}}\lms^2\approx
(2.031\lms)^2\;,\\
\label{rge32c}E_\mathrm{vac}&=&-\frac{3}{16\pi^2}
e^{\frac{17}{6}}\lms^4\approx-0.323\lms^4\;,
\end{eqnarray}
which were obtained upon setting $\omu^2=m^2$ to kill the
logarithms. \\\\We shall now show that, in the $\MSbar$ scheme, the
gap equations (\ref{new8})-(\ref{new9}) have no solution with
$m^2>0$ when the restriction to the horizon is implemented (i.e.
when $\lambda\neq0$). To this purpose, we introduce for $m^2>0$ the
variable
\begin{equation}\label{new13}
t=\frac{\lambda^4}{m^4}\;.
\end{equation}
Evidently, we should only consider $t>0$.\\\\Dividing the gap
equations (\ref{new8})-(\ref{new9}) by $m^2$, they can be rewritten
as\footnote{We have already factored out $m^2$ or $\lambda^3$ since
these are non-zero in the present case.}
\begin{eqnarray}
\label{new14}
\frac{16\pi^2}{g^2N}=\frac{3}{8}\left(-2\ln\frac{m^2}{2\omu^2}+\frac{2}{3}+\frac{1}{\sqrt{1-t}}\ln\frac{t}{\left(1+\sqrt{1-t}\right)^2}-\ln
t\right)\;,
\end{eqnarray}
and
\begin{eqnarray}
\label{new14}-\frac{24}{13}\left(\frac{16\pi^2}{g^2N}\right)+\frac{322}{39}=4\ln\frac{m^2}{2\omu^2}-\frac{4}{3}-\frac{2-t}{\sqrt{1-t}}\ln\frac{t}{\left(1+\sqrt{1-t}\right)^2}+2\ln
t\;,
\end{eqnarray}
where use has been made of the explicit values of $\zeta_0$ and
$\zeta_1$, which can be found in
\cite{Verschelde:2001ia,Browne:2003uv,Dudal:2003by}
\begin{equation}\label{new15}
    \zeta_0=\frac{9}{13}\frac{N^2-1}{N}\;,\;\;\;\;\;\zeta_1=\frac{161}{52}\frac{N^2-1}{16\pi^2}\;,
\end{equation}
The eqns.(\ref{new14})-(\ref{new15}) can be combined to eliminate
$\ln\frac{m^2}{2\omu^2}$, yielding the following condition
\begin{equation}\label{new16}
    \frac{68}{39}\left(\frac{16\pi^2}{g^2N}\right)+\frac{161}{39}=\frac{t}{\sqrt{1-t}}\ln\frac{t}{\left(1+\sqrt{1-t}\right)^2}\equiv F(t)\;.
\end{equation}
It can be checked that $F(t)$ is real-valued and negative for $t>0$,
thus the r.h.s. of eq.(\ref{new16}) is always negative. Since the
l.h.s. of eq.(\ref{new16}) is necessarily positive for a meaningful
result (i.e. $g^2\geq0$), there is no solution with $m^2>0$. As
already mentioned, there are a priori also possible solutions with
$m^2<0$.
\\\\To investigate the existence of a solution with $m^2<0$, it might be instructive to look
again at the gap equations (\ref{new8}) and (\ref{new9}) from
another perspective. We recall that, if the horizon is not
implemented, i.e. $\lambda\equiv0$, the gap equation (\ref{new9})
has two solutions, a perturbative one corresponding to $m^2=0$ (no
condensation) and a non-perturbative one corresponding to the $m^2$
given in eq.(\ref{rge32b}).
\\\\If we momentarily consider $\lambda$ as a free, adjustable parameter of the theory,
eq.(\ref{new9}) dictates how $m^2$ becomes a function of the
parameter $\lambda$. From the result at $\lambda=0$, we could expect
that two branches of solutions would evolve, one starting from the
perturbative and one from the non-perturbative value of $m^2$ at
$\lambda=0$. When $\lambda\equiv0$, the choice for the scale $\omu$
is quite obvious from the requirement that all the logarithms
$\ln\frac{m^2}{\omu^2}$ are vanishing. However, when $\lambda\neq0$,
we notice that there are two kinds of logarithms present, being
$\ln\frac{m^2+\sqrt{m^4-\lambda^4}}{2\omu^2}$ and
$\ln\frac{m^2-\sqrt{m^4-\lambda^4}}{2\omu^2}$. We opt to set
\begin{equation}\label{choiceomu}
    \omu^2=\frac{\left|m^2+\sqrt{m^4-\lambda^4}\right|}{2}\,.
\end{equation}
This reduces to $\omu^2=m^2$ if $\lambda=0$, while it allows
for\footnote{Evidently, $\omu^2$ should be real and positive, hence
the modulus in eq.(\ref{choiceomu}).} $m^2<0$. This is possible if
$m^4\leq\lambda^4$, as it was mentioned below eq.(\ref{10}). In this
case, the size of both logarithms,
$\ln\frac{m^2+\sqrt{m^4-\lambda^4}}{2\omu^2}$ and
$\ln\frac{m^2-\sqrt{m^4-\lambda^4}}{2\omu^2}$, is
determined by their arguments, which are complex conjugate.\\\\
Let us specify to the case $N=3$. In Figure 1, we have plotted the
behaviour of $m^2(\lambda^4)$. We see that next to the
``non-perturbative'' branch of solutions, starting from
$m^2\neq0$, also a ``perturbative'' branch of solutions with
$m^2<0$ is emerging from $m^2=0$, in correspondence with our
expectation.
\begin{figure}[h]\label{fig1}
\begin{center}
  \scalebox{1.4}{\includegraphics{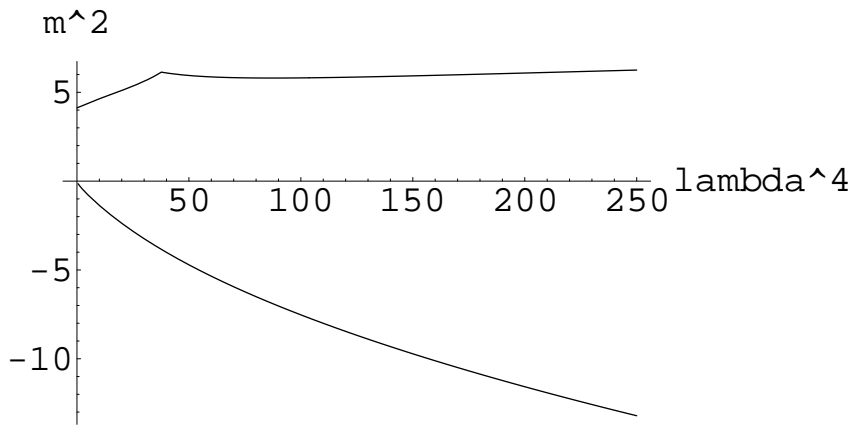}}
\caption{$m^2$ as a function of $\lambda^4$, in units $\lms=1$.}
\end{center}
\end{figure}\\\\
However, $\lambda^4$ is not a free parameter of the theory. We
should require that $\lambda^4$ is such that the doublet
$(\lambda^4,m^2(\lambda^4))$ is a solution of the gap equation
(\ref{new8}), i.e. the horizon condition. In Figure 2, we have
plotted the value of the horizon condition equation, as a function
of $\lambda^4$. It is clear that no solution with $m^2>0$ exists as
the horizon condition never becomes zero. Of course, this is in
correspondence with the foregoing general proof that there is never
such a solution, independently of the choice of $\omu$. However, we
see that there is a single solution with $m^2<0$.
\begin{figure}[h]\label{fig2}
\begin{center}
  \scalebox{1}{\includegraphics{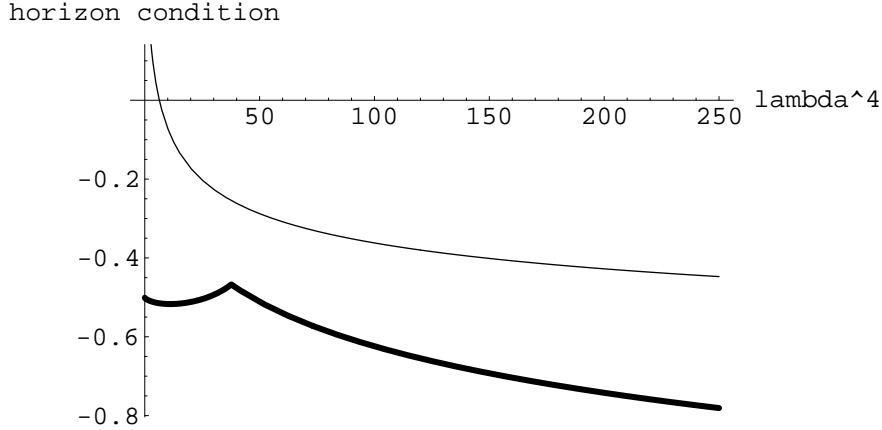}}
\caption{The horizon condition (\ref{new8}) as a function of
$\lambda^4$, in units $\lms=1$. The top curve corresponds to the
solutions of (\ref{new9}) with $m^2<0$ and the lower curve to the
solutions with $m^2>0$.}
\end{center}
\end{figure}\\\\
The corresponding values for the expansion parameter, for the Gribov
and mass parameter, as well as for the vacuum energy are found to be
\begin{eqnarray}
\label{msbaropLa}\frac{g^2N}{16\pi^2}&\approx&1.18\; ,\\
\label{msbaropLb} \lambda^4&\approx&6.351\lms^4\;,\\
\label{msbaropLc}m^2&\approx&
-0.950\lms^2\;,\\
\label{msbaropld}E_\mathrm{vac}&\approx&0.043\lms^4\;, \label{e}
\end{eqnarray}
\subsection{Intermediate comments.}
Although the $\MSbar$ expansion parameter (\ref{msbaropLa}) is too
large to speak about reliable results, we nevertheless would like to
raise some questions. Apparently, the solution of the coupled gap
equations is laying on the ``perturbative'' branch, being the one
with $m^2\leq0$. This gives rise to a positive value for the mass
dimension two gluon condensate $\left\langle A_\mu^2\right\rangle$.
When the restriction on the domain of integration in the path
integral is not implemented, as in the previous papers
\cite{Verschelde:2001ia,Browne:2003uv,Dudal:2003by}, $\left\langle
A_\mu^2\right\rangle$ was necessarily negative, the reason being
that the action should be real-valued, as it was explained below
eq.(\ref{10}). As already explained in the Introduction, a finding a
bit unfortunate is that the vacuum energy is positive, eq.(\ref{e}),
which leads to a negative estimate for the gluon condensate
$\left\langle\frac{g^2}{4\pi^2}F_{\mu\nu}^2\right\rangle$ via the
trace anomaly, eq.(\ref{traceano2}). Essentially, we are thus left
with the following questions:
\begin{itemize}
    \item[(i.)] What is the sign and value of $m^2$ and thus of $\left\langle A_\mu^2\right\rangle$?
    \item[(ii.)] What is the sign and value of $\Evac$ and the
    corresponding value for
    $\left\langle\frac{g^2}{4\pi^2}F_{\mu\nu}^2\right\rangle$?
    \item[(iii.)] Are these values better or not when the operator $A_\mu^2$ is added to
    the original Gribov-Zwanziger model?
\end{itemize}

\sect{Changing and reducing the dependence on the renormalization
scheme.} We have already shown that the vacuum energy obtained in a
one-loop approximation is always positive when the condensation of
the operator $A_\mu^2$ is left out of the discussion, using whatever
renormalization scheme.\\\\To answer the foregoing questions
(i.)-(iii.), one could investigate what happens at two-loop order.
However, due to the already quite complicated structure of the
one-loop effective action and to the fact that the calculations at
higher loop order will not get any easier, this task is beyond the
scope of the present article. Here, we shall mainly focus on the
effects of a change of the renormalization scheme at the one-loop
order. It could happen that, in a scheme different from the $\MSbar$
one, the vacuum energy is negative and/or that the coupling constant
is small enough to speak about trustworthy results, at least
qualitatively.\\\\Since to obtain an optimization of the
renormalization scheme and scale dependence is a rather lengthy
task, we shall not dwell upon technicalities in this section. The
interested reader can find all details in Appendix B. We shall
thus focus on the main results obtained after the optimization.\\\\
Essentially, what we have done is replacing in the effective action
(\ref{new5}) the quantities $m^2$ and $\lambda^4$ by their order by
order renormalization scale and scheme invariant counterparts
$\wm^2$ and $\wl^4$. The residual freedom in the choice of
renormalization scheme can then be reduced to a single parameter
$b_0$, related to coupling constant renormalization. As the vacuum
energy is a physical quantity, it should in principle not depend on
$b_0$. At the same time, the quantities $\wm^2$ and $\wl^4$ should
be $b_0$ independent by construction. This provides one with the
interesting opportunity to fix the redundant parameter $b_0$ by
demanding a minimal dependence on it.\\\\ The final one-loop action
turns out to be given by
\begin{eqnarray}\label{opt12} \Gamma  &=&- \frac{\left(
N^{2}-1\right)}{2N}x^{-2b}\wl
^{4}\left(x+B+(1-2b)\left(\frac{\beta_1}{\beta_0}\ln\frac{x}{\beta_0}-b_0\right)\right)\nonumber\\
&+&\frac{\zeta_0}{2}\wm^4x^{-2a}\left(x+A-\frac{\zeta _{1}}{\zeta
_{0}}+(1-2a)\left(\frac{\beta_1}{\beta_0}\ln\frac{x}{\beta_0}-b_0\right)\right)+\frac{3\left(
N^{2}-1\right) }{256\pi^2}\times
\nonumber\\&&\left[\left(\wm^2x^{-a}+\sqrt{\wm^4x^{-2a}-\wl^4x^{-2b}}\right)^2
\left(\ln\frac{\wm^2x^{-a}+\sqrt{\wm^4x^{-2a}-\wl^4x^{-2b}}}{2\omu^2}-\frac{5}{6}\right)\right.\nonumber\\
&+&\left.\left(\wm^2x^{-a}-\sqrt{\wm^4x^{-2a}-\wl^4x^{-2b}}\right)^2
\left(\ln\frac{\wm^2x^{-a}-\sqrt{\wm^4x^{-2a}-\wl^4x^{-2b}}}{2\omu^2}-\frac{5}{6}\right)\right]
\;,
\end{eqnarray}
while the corresponding gap equations read
\begin{eqnarray}\label{opt13}
\label{opt13} \frac{1}{\wl^3}\frac{\p\Gamma}{\p\wl}&=&0\;,\\
\label{opt13bis}\frac{1}{\wm^2}\frac{\p\Gamma}{\p\wm^2}&=&0\;.
\end{eqnarray}
The definitions of all quantities appearing in the above expressions
can be found in the Appendix B.\\\\ In Figure 3, we collected the
solutions of the scale invariant quantities $\wm^2$ and $\wl^4$ as a
function of $b_0$, while Figure 4 displays the vacuum energy $\Evac$
and the relevant expansion parameter, given by
$y\equiv\frac{N}{16\pi^2 x}$. For completeness, we have also shown
the solutions which correspond to higher values of $\Gamma$ and are
as such describing unstable solutions. These are indicated with the
thinner lines.
\begin{figure}[h]\label{fig8and9}
\begin{tabular}{cc}
  \scalebox{0.95}{\includegraphics{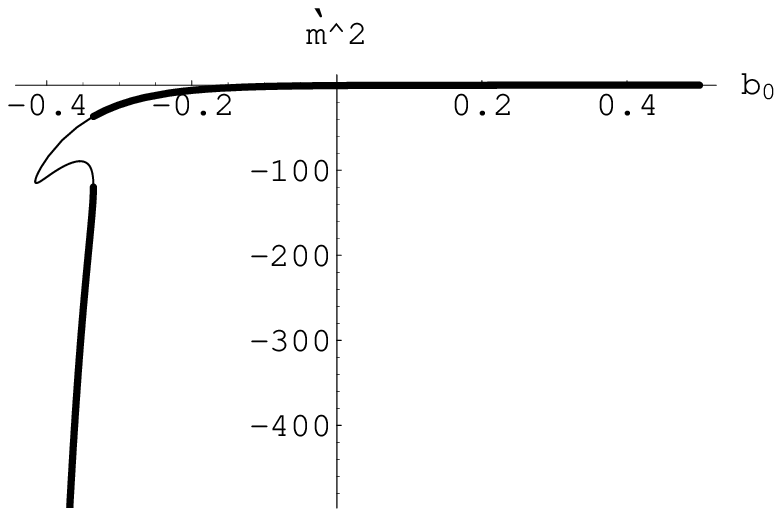}} & \scalebox{0.95}{\includegraphics{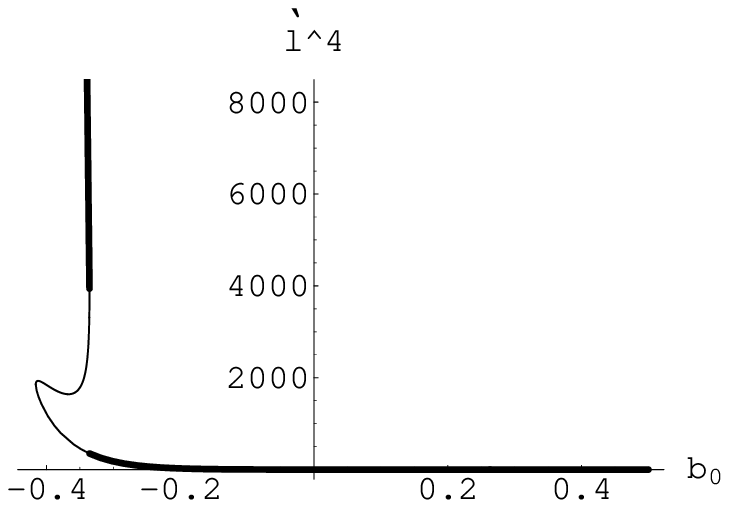}}
  \\
\end{tabular}
\caption{The quantities $\wm^2$ and $\wl^4$ as a function of $b_0$,
in units $\lms=1$.}
\end{figure}

\begin{figure}[h]\label{fig5and7}
\begin{tabular}{cc}
  \scalebox{0.95}{\includegraphics{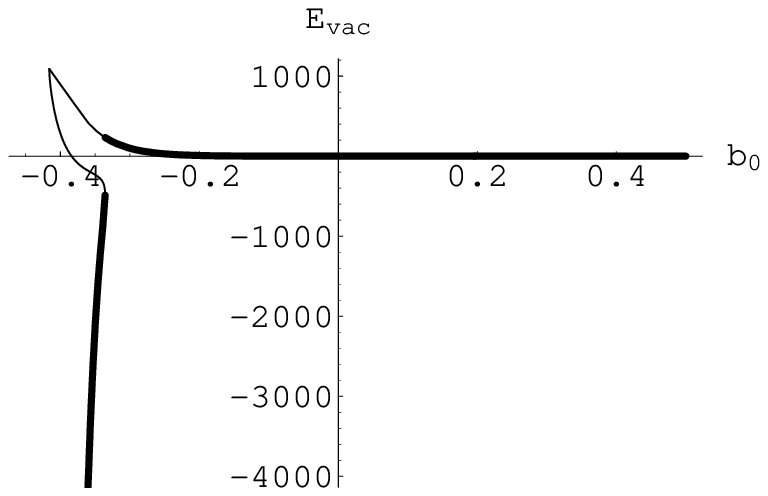}} & \scalebox{0.95}{\includegraphics{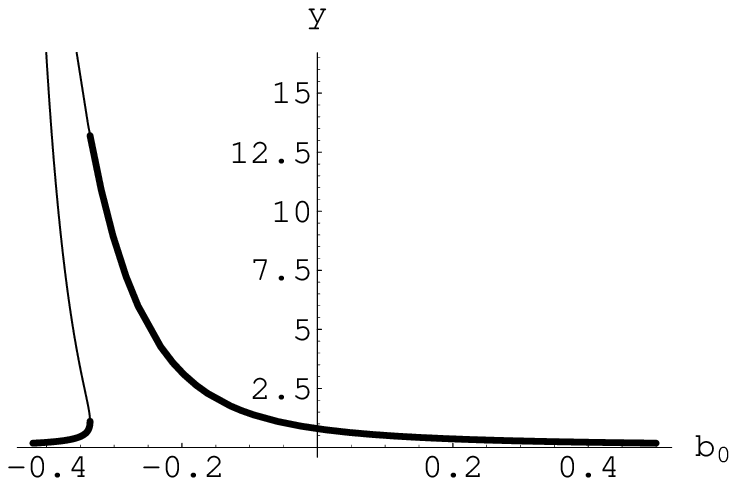}}
  \\
\end{tabular}
\caption{The vacuum energy $\Evac$ and the expansion parameter $y$
as a function of $b_0$, in units $\lms=1$.}
\end{figure}

\subsection{Interpretation of the results.}
Let us first have a look at the plot of vacuum energy, on the l.h.s.
of Figure 4. We notice that for $b_0<-0.33564....$, the vacuum
energy becomes negative. However, we cannot attach any definitive
meaning to this result. In fact, as it can be seen from the Figures
3 and 4, the values of the vacuum energy and the supposedly
minimally $b_0$-dependent quantities $\wm^2$ and $\wl^4$ are
extremely $b_0$-dependent. Very small variations in $b_0$ induce
large fluctuations on e.g. the energy. This is indicative of the
fact that the equations we have solved are not yet stable against
$b_0$-variations in the range of the values obtained for $b_0$. The
behaviour is better for, let us say $b_0>-0.2$. However, in this
case, we find again that the vacuum energy is positive. The vacuum
energy $\Evac$, as well as $\wm^2$ and $\wl^4$ fall of to zero for
growing $b_0$. \\\\ As an example, we set $b_0=0$, which corresponds
to use the $\MSbar$ coupling constant. Then we find, with the
optimized expansion,
\begin{eqnarray}
\label{msbaropLaopt}y&\equiv&\frac{N}{16\pi^2 x}\approx0.796\; ,\\
\label{msbaropLbopt} \wl^4x^{-2b}&\approx&7.939\lms^4\;,\\
\label{msbaropLcopt}\wm^2x^{-a}&\approx&
-0.814\lms^2\;,\\
\label{msbaropldopt}E_\mathrm{vac}&\approx&0.063\lms^4\;,
\end{eqnarray}
results which are in fair agreement with the naive $\MSbar$ results
(\ref{msbaropLa})-(\ref{msbaropld}). We notice that the expansion
parameter $y$ is already smaller than $1$, but still relatively
large, while the vacuum energy is indeed positive.
\\\\ The conclusion than can be drawn from this section is that we
cannot find a reliable result with negative vacuum energy and hence
positive gluon condensate $\left\langle F_{\mu\nu}^2\right\rangle$
using a one-loop approximation. We see therefore that, in order to
be able to give a reasonable answer to the questions concerning the
sign of $m^2$ and $\Evac$ and to get more trustworthy numerical
values, the two-loop evaluation of the effective action $\Gamma$, at
least in the $\MSbar$ scheme, would be very useful.

\sect{Consequences of a non-vanishing Gribov parameter.}
Before
turning to the final conclusions, we shall give in this section a
brief account of some well known consequences stemming from the
presence of the Gribov parameter, to emphasize the important role of
this parameter.

\subsection{The gluon propagator.}
If there is no generation of a mass parameter due to $\left\langle
A_\mu^2\right\rangle$, we can consider just the action (\ref{ll1}).
Then the tree level gluon propagator turns out to be
\begin{equation} \left\langle
A_\mu^aA_\nu^b\right\rangle_p=\delta^{ab}\frac{p^2}{p^4+\frac{\lambda^4}{4}}\left(\delta_{\mu\nu}-
\frac{p_\mu{p}_\nu}{p^2}\right)\;.\label{prop1}
\end{equation}
This result, first pointed out in \cite{Gribov:1977wm}, was obtained
by retaining only the first term of the nonlocal horizon function
(\ref{m3}), corresponding to the approximation
$-\partial{D}\approx-\partial^2$. The gluon propagator,
eq.(\ref{prop1}), is suppressed in the infrared region due to the
presence of the Gribov parameter $\lambda$. In particular, the
presence of this parameters implies that $\left\langle
A_\mu^aA_\nu^b\right\rangle_p$ vanishes at zero momentum, $p=0$.
When the possibility of the existence of a dynamical mass parameter
in the gluon propagator is included, by investigating the
condensation of $A_\mu^2$, the tree level gluon propagator reads
\begin{equation}
\left\langle
A_\mu^aA_\nu^b\right\rangle_p\equiv\delta^{ab}\frac{\mathcal{D}(p^2)}{p^2}\left(\delta_{\mu\nu}-
\frac{p_\mu{p}_\nu}{p^2}\right)=\delta^{ab}\frac{p^2}{p^4+m^2p^2+\frac{\lambda^4}{4}}\left(\delta_{\mu\nu}-
\frac{p_\mu{p}_\nu}{p^2}\right)\;.\label{prop2}
\end{equation}
This type of propagator is sometimes called the Stingl propagator,
from the author who used it as an anzatz for solving the
Schwinger-Dyson equations, see \cite{Stingl:1985hx} for more details
.\\\\However, it should be realized that eq.(\ref{prop2}) describes
only the tree level gluon propagator. In particular, to produce a
plot of the form factor $\mathcal{D}(p^2)$ as a function of the
momentum $p$, which would allow to make a comparison with the
results obtained in lattice simulations, see e.g.
\cite{Bonnet:2001uh} for $N=3$ and
\cite{Langfeld:2001cz,Bloch:2003sk} for $N=2$, one should go beyond
the zeroth order approximation, for example by including higher
order polarization effects and/or trying to perform a
renormalization group improvement. In general, these corrections
will also be dependent on the external momentum $p$.
\subsection{The ghost propagator.}
Even more prominent is the influence of the Gribov parameter on the
infrared behaviour of the ghost propagator, which can be calculated
at one-loop order using the modified gluon propagator (\ref{prop1})
or (\ref{prop2}) with their respective gap equations (\ref{mm5}) and
(\ref{7bis}). In both cases, the infrared behaviour of the ghost
propagator
\cite{Gribov:1977wm,Sobreiro:2004us,Sobreiro:2004yj,Zwanziger:1989mf,Zwanziger:1992qr}
is shown to be
\begin{equation}
\frac{\delta^{ab}}{N^2-1} \left\langle
c^a\oc^b\right\rangle_{p\approx0}\equiv\left.\frac{1}{p^2}\mathcal{G}(p^2)\right|_{p\approx0}\approx\frac{4}{3Ng^2\mathcal{J}p^4}\;,\label{prop3}
\end{equation}
where $\mathcal{J}$ stands for the real, finite integral given by
\begin{equation}
\mathcal{J}=\int\frac{d^4k}{(2\pi)^4}\frac{1}{k^2\left(k^4+m^2k^2+\frac{\lambda^4}{4}\right)}\;.\label{int}
\end{equation}
The original Gribov-Zwanziger model corresponds to $m^2\equiv0$.
Thus, the ghost propagator is strongly enhanced in the infrared
region compared to the perturbative behaviour, if the restriction to
the first Gribov region is taken into account. It is important to
notice that this behaviour of the ghost propagator is preserved in
the present treatment, due to the peculiar form of the gap equation
(\ref{7bis}) implementing the horizon condition. In particular, from
the expression for the effective action in eq.(\ref{7}), one sees
that, while the term quadratic in the field $\sigma$, i.e.
$\frac{\sigma^2}{2g^2\zeta}$, contains the LCO parameter $\zeta$,
the first term which depends on the Gribov parameter, i.e.
$-4(N^2-1)\gamma^4$, does not contain any such new LCO parameter.
This important feature follows from the fact that no new parameter
has to be introduced in order to renormalize the term $\left( M_{\mu
}^{ai}V_{\mu }^{ai}-U_{\mu }^{ai}N_{\mu }^{ai}\right)$, as remarked
in eq.(\ref{co7}). While the parameter $\zeta$ is required to take
into account the ultraviolet divergences of the vacuum correlator
$\langle A^2_\mu(x) A^2_\nu(y) \rangle$, which are proportional to
$\tau^2$, no such a parameter is needed for $\left( M_{\mu
}^{ai}V_{\mu }^{ai}-U_{\mu }^{ai}N_{\mu }^{ai}\right)$ which, upon
setting the external sources to their physical values, gives rise to
term $-4(N^2-1)\gamma^4$ in the expression (\ref{7}). Said
otherwise, this term is not affected by the presence of a new
parameter which would be required if eq.(\ref{co7}) would not hold.
As a consequence, the factor ``1'' appearing in the left hand side
of the gap equation (\ref{7bis}) is, so to speak, left unchanged by
the quantum corrections. It is precisely that property which
ensures, through a delicate cancelation mechanism, see
\cite{Gribov:1977wm,Sobreiro:2004us,Zwanziger:1989mf,Zwanziger:1992qr},
the infrared enhancement of the ghost propagator.
\\\\Analogously to the case of the gluon propagator,
a more detailed study of higher order corrections would be needed in
order to obtain a plot of the ghost form factor $\mathcal{G}(p^2)$.

\subsection{The strong coupling constant.}
Usually, a nonperturbative definition of the renormalized strong
coupling constant $\alpha_R$ can be written down from the knowledge
of the gluon and ghost propagators as, see e.g.
\cite{vonSmekal:1997is,Bloch:2003sk}
\begin{equation}
\alpha_R(p^2)=\alpha_R(\mu)\mathcal{D}(p^2,\mu)\mathcal{G}^2(p^2,\mu)\;,\label{coup}
\end{equation}
where $\mathcal{D}$ and $\mathcal{G}$ stand for the gluon and ghost
form factors as defined before. This definition represents a kind of
nonperturbative extension of the perturbative results (\ref{co5}).
According to Schwinger-Dyson studies
\cite{Atkinson:1997tu,Atkinson:1998zc,Alkofer:2000wg,Watson:2001yv,Zwanziger:2001kw,Lerche:2002ep},
those form factors satisfy a power law behaviour in the infrared
\begin{eqnarray}
\lim_{p\rightarrow0}\mathcal{D}(p^2)&\propto&\left(p^2\right)^\theta\;,
\nonumber\\
\lim_{p\rightarrow0}\mathcal{G}(p^2)&\propto&\left(p^2\right)^\omega\;,\label{form2}
\end{eqnarray}
where the infrared exponents $\theta$ and $\omega$ obey the sum rule
\begin{equation}
\theta+2\omega=0\;.\label{sum}
\end{equation}
Such a sum rule suggests the development of an infrared fixed point
for the renormalized coupling constant, (\ref{coup}), as also
pointed out by lattice simulations for the $SU(2)$ as well as for
the $SU(3)$ case \cite{Bloch:2003sk,Furui:2003jr,Furui:2004cx},
\begin{equation}
\lim_{p\rightarrow0}\alpha(p^2)=\alpha_c\;.\label{crit}
\end{equation}
The existence of a fixed point in this reasoning is dependent on the
sum rule rather than on the precise value of the exponents. We refer
to the already quoted literature for more details on the value of
these exponents. We end by noticing that the form factors of the
gluon and ghost propagator in our zeroth order approximation give
rise to the sum rule (\ref{sum}), since we have $\theta=2$ and
$\omega=-1$. Moreover, without Gribov parameter, the sum rule
(\ref{sum}) is lost, and thus there is no indication for an infrared
fixed point.
\subsection{Positivity violation.}
The behaviour of the gluon propagator is sometimes used as an
\emph{indication} of confinement of gluons by means of the so called
positivity violation, see e.g.
\cite{Alkofer:2003jj,Cucchieri:2004mf} and references therein.
\\\\Briefly, when the Euclidean gluon propagator
$D(p)\equiv\frac{\mathcal{D}(p^2)}{p^2}$ is written through a
spectral representation as
\begin{equation}\label{spectr1}
    D(p)=\int_0^{+\infty}dM^2\frac{\rho(M^2)}{p^2+M^2}\;,
\end{equation}
the spectral density $\rho(M^2)$ should be positive in order to have
a K\"{a}llen-Lehmann representation, making possible the
interpretation of the fields in term of stable particles. We refer
to \cite{Alkofer:2003jj,Cucchieri:2004mf} for more details. One can
define the temporal correlator \cite{Cucchieri:2004mf}
\begin{equation}\label{spectr2}
    \mathcal{C}(t)=\int_0^{+\infty}dM\rho(M^2)e^{-Mt}\;,
\end{equation}
which is certainly positive for positive $\rho(M^2)$. The inverse is
not necessarily true. $\mathcal{C}(t)$ can be also positive for a
$\rho(M^2)$ attaining negative values. However, if $\mathcal{C}(t)$
becomes negative for certain $t$, then a fortiori $\rho(M^2)$ cannot
be always positive. Using a contour integration argument, it is not
difficult to show that $\mathcal{C}(t)$ can be rewritten as
\begin{equation}\label{spectr3}
    \mathcal{C}(t)=\frac{1}{2\pi}\int_{-\infty}^{+\infty}e^{-ipt}D(p)dp\;.
\end{equation}
Let us consider the function $\mathcal{C}(t)$ using the tree level
propagator (\ref{prop2}), thus using
\begin{equation}\label{spectr4}
    D(p)=\frac{p^2}{p^4+p^2m^2+\frac{\lambda^4}{4}}\;.
\end{equation}
We can consider several cases\footnote{Each of the following
expressions for $\mathcal{C}(t)$ is obtainable via contour
integration.}:
\begin{itemize}
    \item if $\lambda=0$ (thus $m^2>0$), one shall find that
\begin{equation}\label{ct2}
    \mathcal{C}(t)=\frac{e^{-mt}}{2m}\;.
\end{equation}
This function is always positive.
    \item if $m^2=0$,
    \begin{equation}\label{ct1}
    \mathcal{C}(t)=\frac{e^{-\frac{Lt}{2}}}{2L}\left(\cos\frac{Lt}{2}-\sin\frac{Lt}{2}\right)\;,
    \end{equation}
and clearly, this function will attain negative values for certain
$t$.
    \item in any other case, the correlator $\mathcal{C}(t)$ is
    found to be
\begin{equation}\label{ct3}
    \mathcal{C}(t)=\frac{1}{2}\left[\frac{\sqrt{\omega_1}}{\omega_1-\omega_2}e^{-\sqrt{\omega_1}t}
    +\frac{\sqrt{\omega_2}}{\omega_2-\omega_1}e^{-\sqrt{\omega_2}t}\right]
\end{equation}
where the decomposition
\begin{equation}\label{rippedgluon}
    \frac{p^2}{p^4+p^2m^2+\frac{\lambda^4}{4}}=\frac{\omega_1}{\omega_1-\omega_2}\frac{1}{p^2+\omega_1}-\frac{\omega_2}{\omega_1-\omega_2}\frac{1}{p^2+\omega_2}\;,
\end{equation}
has been employed. It is understood that $\sqrt{\omega_1}$
($\sqrt{\omega_2}$) is the root having a positive real part. \\\\If
we assume that $\wm^4>\lambda^4$, then $\omega_1>\omega_2$ and
$\mathcal{C}(t)$ becomes negative for
$t>\frac{1}{2}\frac{\ln\frac{\omega_1}{\omega_2}}{\omega_1-\omega_2}$.
In the case that $\wm^4=\lambda^4$, or $\omega_1=\omega_2$, one
finds that
$\mathcal{C}(t)=\frac{e^{-\sqrt{\omega_1}t}}{4\sqrt{\omega_1}}(1-\sqrt{\omega_1}t)$,
which can also become negative. If $\wm^4<\lambda^4$, we can
reintroduce the complex polar coordinates $R$ and $\phi$ for the
complex conjugate quantities $\omega_1$ and $\omega_2$. If
$\cos\frac{\phi}{2}\geq0$, eq.(\ref{ct3}) can be rewritten as
\begin{equation}\label{ct3bis}
    \mathcal{C}(t)=\frac{1}{2\sqrt{R}\sin\phi}e^{-\sqrt{R}\cos\left(\frac{\phi}{2}\right)t}\sin\left(\frac{\phi}{2}-\sqrt{R}\sin\left(\frac{\phi}{2}\right)t\right)
\end{equation}
By choosing an appropriate value of $t>0$, also this expression can
be made negative. An analogous expression and conclusion can be
derived in case that $\cos\frac{\phi}{2}<0$
\end{itemize}
We conclude that, when the restriction to the Gribov region $\Omega$
is implemented, the function $\mathcal{C}(t)$ exhibits a violation
of positivity when the tree level propagator is used, with our
without the inclusion of $\left\langle A_\mu^2\right\rangle$.
\\\\The goal of this section was merely to provide some interesting
consequences when the restriction to the first Gribov region
$\Omega$ is implemented. Higher loop effects, which shall be
momentum dependent, would also influence the behaviour of the gluon
and ghost propagator. Hence, to give a sensible interpretation of
the behaviour of e.g. the form factors and of the strong coupling
constant $\alpha_R$, a more detailed analysis than a tree level one
is necessary. This is however far beyond the aim of this work.

\sect{Conclusion.} In this work we have considered $SU(N)$ Euclidean
Yang-Mills theories in the Landau gauge, $\p_\mu A_\mu=0$. We have
studied the condensation of the dimension two composite operator
$A_\mu^2$ when the restriction to the Gribov region $\Omega$ is
taken into account. Such a restriction is needed due to the presence
of the Gribov copies \cite{Gribov:1977wm}, which are known to affect
the Landau gauge. \\\\
In a previous work \cite{Sobreiro:2004us}, the consequences of the
restriction to the region $\Omega$ in the presence of a dynamical
mass parameter due to the gluon condensate $\left\langle A_\mu^a
A_\mu^a \right\rangle$ were studied by following Gribov's seminal
work \cite{Gribov:1977wm}. Here, we have relied on Zwanziger's
action \cite{Zwanziger:1989mf,Zwanziger:1992qr}, which allows to
implement the restriction to the Gribov region $\Omega$ within a
local and renormalizable framework. We have been able to show that
Zwanziger's action remains renormalizable to all orders of
perturbation theory in the presence of the operator $A_\mu^2$,
introduced through the local composite operator technique
\cite{Verschelde:2001ia,Browne:2003uv,Knecht:2001cc,Dudal:2003by}.
The effective action, constructed via the local composite operator
formalism \cite{Verschelde:2001ia} obeys a homogeneous
renormalization group. The explicit form of the one-loop effective
action has been worked out. We have seen that, considering the
original Gribov-Zwanziger model, i.e. without including the operator
$A_\mu^2$, the vacuum energy is always positive at one-loop order,
independently from the choice of the renormalization scheme. A
positive vacuum energy would give rise to a negative value for the
gauge invariant gluon condensate $\left\langle
F_{\mu\nu}^2\right\rangle$, through the trace anomaly. Furthermore,
by adding the operator $A_\mu^2$, we have proven that there is no
solution of the two coupled gap equations at the one-loop order in
the $\MSbar$ scheme with $\left\langle A_\mu^2\right\rangle<0$.
Nevertheless, when $\left\langle A_\mu^2\right\rangle>0$, a solution
of the gap equations was found, although the corresponding expansion
parameter was too large and the vacuum energy still positive.
\\\\In order to find out what happens in other schemes,
we performed a detailed study, at lowest order, of the influence
of the renormalization scheme. We have been able to reduce the
freedom of the choice of the renormalization scheme to two
parameters, namely the renormalization scale $\omu$ and a
parameter $b_0$, associated to the coupling constant
renormalization. We reexpressed the effective action in terms of
the mass parameter $\wm$ and Gribov parameter $\wl$, which are
renormalization scheme and scale independent order by order. The
resulting gap equations for these parameters have been solved
numerically. Although a solution with negative vacuum energy was
found, we have been unable to attach any definitive meaning to it.
This is due to the fact that the results obtained turn out to be
strongly dependent from the parameter $b_0$. This brought us to
the conclusion that we should extend our calculations to a higher
order to obtain more sensible numerical estimates.
\\\\ The mass parameters $\wm$ and $\wl$ are of a
nonperturbative nature and appear in the gluon and ghost propagator.
Even if we lack reliable estimates for these parameters, some
already known interesting features can be recovered. For a nonzero
mass and Gribov parameter, there is a \emph{qualitative} agreement
with the behaviour found in lattice simulations and Schwinger-Dyson
studies: a suppressed gluon and enhanced ghost propagator in the
infrared, while further consequences of the Gribov parameter are
e.g. the possible existence of an infrared fixed point for the
strong coupling constant and the violation of positivity related to
the gluon propagator.

\section*{Acknowledgments.} The Conselho Nacional de Desenvolvimento
Cient\'{i}fico e Tecnol\'{o}gico (CNPq-Brazil), the Faperj, Funda{\c
c}{\~a}o de Amparo {\`a} Pesquisa do Estado do Rio de Janeiro,
the SR2-UERJ and the Coordena{\c{c}}{\~{a}}o de Aperfei{\c{c}}%
oamento de Pessoal de N{\'{i}}vel Superior (CAPES) are gratefully
acknowledged for financial support. D.~Dudal would like to
acknowledge the warm hospitality at the UERJ, where part of this
work was done, while R.~F.~Sobreiro would like to acknowledge the
kind hospitality at the Department of Mathematical Physics and
Astronomy of the Ghent University, where this work was completed.

\appendix
\section{Appendix A.}
In this Appendix, we have collected all details of the
multiplicative renormalization of the Zwanziger action in the
presence of the operator $A_\mu^2$.
\subsection{Ward identities.}
In order to begin with the algebraic characterization of the most
general counterterm needed for the renormalizability of the complete
action $\Sigma $ of eq.(\ref{r1}), let us first give the set of Ward
identities which are fulfilled by $\Sigma $. These are
\begin{itemize}
\item  the Slavnov-Taylor identity
\begin{equation}
\mathcal{S}(\Sigma )=0\;,  \label{r9}
\end{equation}
with
\begin{eqnarray}
\mathcal{S}(\Sigma ) &=&\int d^{4}x\left( \frac{\delta \Sigma
}{\delta K_{\mu }^{a}}\frac{\delta \Sigma }{\delta A_{\mu
}^{a}}+\frac{\delta \Sigma }{\delta L^{a}}\frac{\delta \Sigma
}{\delta c^{a}}+b^{a}\frac{\delta \Sigma
}{\delta \overline{c}^{a}}+\overline{\varphi }_{i}^{a}\frac{\delta \Sigma }{%
\delta \overline{\omega }_{i}^{a}}+\omega _{i}^{a}\frac{\delta \Sigma }{%
\delta \varphi _{i}^{a}}\right.   \nonumber \\
&\;&\;\;\;\;\;\;\;\;\;\;\;\;+\left.M_{\mu }^{ai}\frac{\delta \Sigma
}{\delta U_{\mu
}^{ai}}+N_{\mu }^{ai}\frac{\delta \Sigma }{\delta V_{\mu }^{ai}}+\tau \frac{%
\delta \Sigma }{\delta \eta }\right) \;,  \label{r10}
\end{eqnarray}

\item  the Landau gauge condition and the antighost equation
\begin{eqnarray}
\frac{\delta \Sigma }{\delta b^{a}}&=&\partial_\mu A_\mu^{a}\;,
\label{r11}\\ \frac{\delta \Sigma }{\delta
\overline{c}^{a}}+\partial _{\mu }\frac{\delta \Sigma }{\delta
K_{\mu }^{a}}&=&0\;,  \label{r12}
\end{eqnarray}

\item  the ghost Ward identity
\begin{equation}
\mathcal{G}^{a}\Sigma =\Delta _{\mathrm{cl}}^{a}\;,  \label{r13}
\end{equation}
with
\begin{eqnarray}
\mathcal{G}^{a} &=&\int d^{4}x\left( \frac{\delta }{\delta c^{a}}%
+gf^{abc}\left( \overline{c}^{b}\frac{\delta }{\delta b^{c}}+\varphi _{i}^{b}%
\frac{\delta }{\delta \omega _{i}^{c}}+\overline{\omega }_{i}^{b}\frac{%
\delta }{\delta \overline{\varphi }_{i}^{c}}+V_{\mu }^{bi}\frac{\delta }{%
\delta N_{\mu }^{ci}}+U_{\mu }^{bi}\frac{\delta }{\delta M_{\mu }^{ci}}%
\right) \right) \;,  \nonumber \\
&&  \label{r14}
\end{eqnarray}
and
\begin{equation}
\Delta _{\mathrm{cl}}^{a}=g\int d^{4}xf^{abc}\left( K_{\mu
}^{b}A_{\mu }^{c}-L^{b}c^{c}\right) \;.  \label{r15}
\end{equation}
Notice that the term $\Delta _{\mathrm{cl}}^{a}$, being linear in
the quantum fields $A_{\mu }^{a}$, $c^{a}$, is a classical breaking.

\item  the linearly broken local constraints
\begin{equation}
\frac{\delta \Sigma }{\delta \overline{\varphi }^{ai}}+\partial _{\mu }\frac{%
\delta \Sigma }{\delta M_{\mu }^{ai}}=gf^{abc}A_{\mu }^{b}V_{\mu
}^{ci}\;, \label{r16}
\end{equation}
\begin{equation}
\frac{\delta \Sigma }{\delta \omega ^{ai}}+\partial _{\mu
}\frac{\delta \Sigma }{\delta N_{\mu
}^{ai}}-gf^{abc}\overline{\omega }^{bi}\frac{\delta \Sigma }{\delta
b^{c}}=gf^{abc}A_{\mu }^{b}U_{\mu }^{ci}\;,
\label{r17}\\
\end{equation}
\begin{equation}
\frac{\delta \Sigma }{\delta \overline{\omega }^{ai}}+\partial _{\mu }\frac{%
\delta \Sigma }{\delta U_{\mu }^{ai}}-gf^{abc}V_{\mu
}^{bi}\frac{\delta \Sigma }{\delta K_{\mu }^{c}}=-gf^{abc}A_{\mu
}^{b}N_{\mu }^{ci} \;, \label{r18}\\
\end{equation}
\begin{equation}
\frac{\delta \Sigma }{\delta \varphi ^{ai}}+\partial _{\mu
}\frac{\delta \Sigma }{\delta V_{\mu
}^{ai}}-gf^{abc}\overline{\varphi }^{bi}\frac{\delta
\Sigma }{\delta b^{c}}-gf^{abc}\overline{\omega }^{bi}\frac{\delta \Sigma }{%
\delta \overline{c}^{c}}-gf^{abc}U_{\mu }^{bi}\frac{\delta \Sigma
}{\delta K_{\mu }^{c}} =gf^{abc}A_{\mu }^{b}M_{\mu }^{ci} \;,
\label{r19}
\end{equation}

\item  the integrated Ward identity
\begin{equation}
\int d^{4}x\left( c^{a}\frac{\delta \Sigma }{\delta \omega ^{ai}}+\overline{%
\omega }^{ai}\frac{\delta \Sigma }{\delta \overline{c}^{a}}+U_{\mu }^{ai}%
\frac{\delta \Sigma }{\delta K_{\mu }^{a}}\right) =0\;,  \label{r20}
\end{equation}

\item  the exact $\mathcal{R}_{ij}$ symmetry
\begin{equation}
\mathcal{R}_{ij}\Sigma =0\;,  \label{r21}
\end{equation}
with
\begin{equation}
\mathcal{R}_{ij}=\int d^{4}x\left( \varphi _{i}^{a}\frac{\delta
}{\delta
\omega _{j}^{a}}-\overline{\omega }_{j}^{a}\frac{\delta }{\delta \overline{%
\varphi }_{i}^{a}}+V_{\mu }^{ai}\frac{\delta }{\delta N_{\mu
}^{ai}}-U_{\mu }^{ai}\frac{\delta }{\delta M_{\mu }^{ai}}\right) \;.
\label{r22}
\end{equation}
\end{itemize}

\subsection{Algebraic characterization of the counterterm.}
Having established all the Ward identities fulfilled by the complete action $%
\Sigma $, we can now turn to the characterization of the most
general allowed counterterm $\Sigma ^{c}$. Following the algebraic
renormalization procedure \cite{Piguet:1995er}, $\Sigma ^{c}$ is an
integrated local polynomial in the fields and sources with dimension
bounded by four, with vanishing ghost number and $Q_{f}$-charge,
obeying the following constraints
\begin{eqnarray}
\frac{\delta \Sigma ^{c}}{\delta \varphi ^{ai}}+\partial _{\mu
}\frac{\delta
\Sigma ^{c}}{\delta V_{\mu }^{ai}}-gf^{abc}\overline{\omega }^{bi}\frac{%
\delta \Sigma ^{c}}{\delta \overline{c}^{c}}-gf^{abc}U_{\mu }^{bi}\frac{%
\delta \Sigma ^{c}}{\delta K_{\mu }^{c}} &=&0\;,  \nonumber \\
\frac{\delta \Sigma ^{c}}{\delta \overline{\omega }^{ai}}+\partial _{\mu }%
\frac{\delta \Sigma ^{c}}{\delta U_{\mu }^{ai}}-gf^{abc}V_{\mu }^{bi}\frac{%
\delta \Sigma ^{c}}{\delta K_{\mu }^{c}} &=&0\;,  \nonumber \\
\frac{\delta \Sigma ^{c}}{\delta \omega ^{ai}}+\partial _{\mu
}\frac{\delta \Sigma ^{c}}{\delta N_{\mu }^{ai}} &=&0\;,
\nonumber\\
\frac{\delta \Sigma }{\delta \overline{\varphi }^{ai}}+\partial _{\mu }\frac{%
\delta \Sigma }{\delta M_{\mu }^{ai}} &=&0\;,  \nonumber
\end{eqnarray}
\begin{eqnarray}
\frac{\delta \Sigma }{\delta \overline{c}^{a}}+\partial _{\mu
}\frac{\delta
\Sigma }{\delta K_{\mu }^{a}} &=&0\;,  \nonumber \\
\frac{\delta \Sigma ^{c}}{\delta b^{a}} &=&0\;,  \label{c2}
\end{eqnarray}
\begin{equation}
\mathcal{G}^{a}\Sigma ^{c}=0\,,  \label{c3}
\end{equation}
\begin{equation}
\int d^{4}x\left( c^{a}\frac{\delta \Sigma ^{c}}{\delta \omega ^{ai}}+%
\overline{\omega }^{ai}\frac{\delta \Sigma ^{c}}{\delta \overline{c}^{a}}%
+U_{\mu }^{ai}\frac{\delta \Sigma ^{c}}{\delta K_{\mu }^{a}}\right)
=0\;, \label{c4}
\end{equation}
\begin{equation}
\mathcal{R}_{ij}\Sigma ^{c}=0\;,  \label{c5}
\end{equation}
and
\begin{equation}
\mathcal{B}_{\Sigma }\Sigma ^{c}=0\;,  \label{c6}
\end{equation}
where $\mathcal{B}_{\Sigma }$ is the nilpotent linearized
Slavnov-Taylor operator
\begin{eqnarray}
\mathcal{B}_{\Sigma } &=&\int d^{4}x\left( \frac{\delta \Sigma
}{\delta
K_{\mu }^{a}}\frac{\delta }{\delta A_{\mu }^{a}}+\frac{\delta \Sigma }{%
\delta A_{\mu }^{a}}\frac{\delta }{\delta K_{\mu }^{a}}+\frac{\delta
\Sigma }{\delta L^{a}}\frac{\delta }{\delta c^{a}}+\frac{\delta
\Sigma }{\delta
c^{a}}\frac{\delta }{\delta L^{a}}+b^{a}\frac{\delta }{\delta \overline{c}%
^{a}}\right.  \nonumber \\
&\;&\;\;\;\;\;\;\;\;\;\;\;\;+\left. \overline{\varphi
}_{i}^{a}\frac{\delta }{\delta \overline{\omega }_{i}^{a}}+\omega
_{i}^{a}\frac{\delta }{\delta \varphi
_{i}^{a}}+M_{\mu }^{ai}\frac{\delta }{\delta U_{\mu }^{ai}}+N_{\mu }^{ai}%
\frac{\delta }{\delta V_{\mu }^{ai}}+\tau \frac{\delta }{\delta \eta }%
\right) \;,  \label{c7}
\end{eqnarray}
\begin{equation}
\mathcal{B}_{\Sigma }\mathcal{B}_{\Sigma }=0\;.  \label{c8}
\end{equation}
As it was shown in
\cite{Zwanziger:1989mf,Zwanziger:1992qr,Maggiore:1993wq}, the
constraints (\ref{c2}) imply that $\Sigma ^{c}$ does not depend on
the Lagrange multiplier $b^{a}$, and that the antighost
$\overline{c}^{a}$ and the $i$-valued fields $\varphi
_{i}^{a}$, $\omega _{i}^{a}$, $\overline{\varphi }_{i}^{a}$, $\overline{%
\omega }_{i}^{a}$ can enter only through the combinations
\begin{eqnarray}
\widetilde{K}_{\mu }^{a} &=&K_{\mu }^{a}+\partial _{\mu }\overline{c}%
^{a}-gf^{abc}\widetilde{U}_{\mu }^{bi}\varphi ^{ci}-gf^{abc}V_{\mu }^{bi}%
\overline{\omega }^{ci}\;,  \nonumber \\
\widetilde{U}_{\mu }^{ai} &=&U_{\mu }^{ai}+\partial _{\mu }\overline{\omega }%
^{ai}\;,  \nonumber \\
\widetilde{V}_{\mu }^{ai} &=&V_{\mu }^{ai}+\partial _{\mu }\varphi
^{ai}\;,
\nonumber \\
\widetilde{N}_{\mu }^{ai} &=&N_{\mu }^{ai}+\partial _{\mu }\omega
^{ai}\;,
\nonumber \\
\widetilde{M}_{\mu }^{ai} &=&V_{\mu }^{ai}+\partial _{\mu
}\overline{\varphi }^{ai}\;.  \label{c9}
\end{eqnarray}
Therefore, $\Sigma ^{c}$ can be parametrized as follows
\begin{eqnarray}
\Sigma ^{c} &=&S^{c}(A)+\int d^{4}x\left( a_{1}gf^{abc}L^{a}c^{b}c^{c}+a_{2}%
\widetilde{K}_{\mu }^{a}\partial _{\mu }c^{a}+a_{3}gf^{abc}\widetilde{K}%
_{\mu }^{a}A_{\mu }^{b}c^{c}+a_{4}f^{abc}\widetilde{V}_{\mu }^{ai}\widetilde{%
U}_{\mu }^{bi}c^{c}\right.  \nonumber \\
&+&\left.a_{5}\widetilde{V}_{\mu }^{ai}\widetilde{M}_{\mu
}^{ai}+a_{6}\widetilde{U}_{\mu }^{ai}\widetilde{N}_{\mu }^{ai}+\frac{a_{7}}{2%
}\tau A_{\mu }^{a}A_{\mu }^{a}+\frac{a_{8}}{2}\zeta \tau
^{2}+a_{9}\eta A_{\mu }^{a}\partial _{\mu }c^{a}+a_{10}\eta
c^{a}\partial A^{a}\right) \;,  \label{c10}
\end{eqnarray}
where $S^{c}(A)$ depends only on the gauge field $A_{\mu }^{a}$, and with $%
a_{1}$, ..., $a_{10}$ arbitrary parameters. Notice, however, that
there is
no mixing in expression (\ref{c10}) between $\widetilde{M}%
_{\mu }^{ai}$, $\widetilde{N}_{\mu }^{ai}$, $\widetilde{V}_{\mu }^{ai}$, $%
\widetilde{U}_{\mu }^{ai}$ and the sources $\tau $, $\eta $. This is
due to the dimensionality and to the $Q_{f}$-charge. It is precisely
the absence of this mixing that will ensure the renormalizability of
the Zwanziger action in the presence of the composite operator
$A_{\mu }^{a}A_{\mu }^{a}$. From the ghost equation (\ref{c3}) it
follows
\begin{eqnarray}
a_{1} &=&a_{3}=a_{10}=0\;,  \nonumber \\
a_{4} &=&-g(a_{6}+a_{5})\;.  \label{c11}
\end{eqnarray}
>From the equations (\ref{c4}) and (\ref{c5}) we obtain
\begin{equation}
a_{6}=-a_{2}\;.  \label{c12}
\end{equation}
Finally, from eq.(\ref{c6}) it turns out that
\begin{eqnarray}
a_{5} &=&a_{2}\;,  \nonumber \\
a_{9} &=&a_{7}-a_{2}\;,  \label{c13}
\end{eqnarray}
and
\begin{equation}
S^{c}(A)=a_{0}S_{YM}\;+a_{2}\int d^{4}xA_{\mu }^{a}\frac{\delta S_{YM}}{%
\delta A_{\mu }^{a}}\;.  \label{c14}
\end{equation}
In summary, the most general local invariant counterterm compatible
with all
Ward identities contains four arbitrary parameters, $a_{0}$, $a_{2}$, $a_{7}$%
, $a_{8}$, and reads
\begin{eqnarray}
\Sigma ^{c} =a_{0}S_{YM}&+&a_{2}\int d^{4}x\left( A_{\mu
}^{a}\frac{\delta
S_{YM}}{\delta A_{\mu }^{a}}+\widetilde{K}_{\mu }^{a}\partial _{\mu }c^{a}+%
\widetilde{V}_{\mu }^{ai}\widetilde{M}_{\mu }^{ai}-\widetilde{U}_{\mu }^{ai}%
\widetilde{N}_{\mu }^{ai}\right)  \nonumber \\
&+&\int d^{4}x\left( \frac{a_{7}}{2}\tau A_{\mu }^{a}A_{\mu }^{a}+%
\frac{a_{8}}{2}\zeta \tau ^{2}+\left( a_{7}-a_{2}\right) \eta A_{\mu
}^{a}\partial _{\mu }c^{a}\right) \;.  \label{c15}
\end{eqnarray}

\subsection{Stability and renormalization constants.}
Having determined the most general local invariant counterterm
$\Sigma ^{c}$ compatible with all Ward identities, it remains to
check that the starting action $\Sigma $ is stable, i.e. that
$\Sigma ^{c}$ can be reabsorbed through the renormalization of the
parameters, fields and sources of $\Sigma $. According to expression
(\ref{c15}), $\Sigma ^{c}$ contains four arbitrary parameters
$a_{0}$, $a_{2}$, $a_{7}$, $a_{8}$, which correspond in fact to a
multiplicative renormalization of the gauge coupling constant $g$,
the parameters $\zeta $, and of the fields $\phi
=(A_{\mu }^{a}$, $c^{a}$, $\overline{c}^{a}$, $b^{a}$, $\varphi _{i}^{a}$, $%
\omega _{i}^{a}$, $\overline{\varphi }_{i}^{a}$, $\overline{\omega }%
_{i}^{a}) $ and sources $\Phi =(K^{a\mu }$, $L^{a}$, $M_{\mu }^{ai}$, $%
N_{\mu }^{ai}$, $V_{\mu }^{ai}$, $U_{\mu }^{ai},\tau $, $\eta )$,
according to
\begin{equation}
\Sigma (g,\zeta ,\phi ,\Phi )+\eta \Sigma ^{c}=\Sigma (g_{o},\zeta
_{o},\phi _{o},\Phi _{o})+O(\eta ^{2})\;, \label{co1}
\end{equation}
with
\begin{equation}
g_{o}=Z_{g}g\;,\;\;\;\;\zeta _{o}=Z_{\zeta }\zeta \;,\;\;\;\;
\label{co2}
\end{equation}
and
\begin{eqnarray}
\phi _{o} &=&Z_{\phi }^{1/2}\phi \;,\;\;\;  \nonumber \\
\Phi _{o}\; &=&Z_{\Phi }\Phi \;.  \label{co3}
\end{eqnarray}
The coefficients $a_{0}$, $a_{2}$ are easily seen to be related to
the
renormalization of the gauge coupling constant $g$ and of the gauge field $%
A_{\mu }^{a}$,
\begin{eqnarray}
Z_{g} &=&\left( 1+\eta \frac{a_{0}}{2}\right) \;,  \nonumber \\
Z_{A}^{1/2} &=&\left( 1+\eta \left( a_{2}-\frac{a_{0}}{2}\right)
\right) \;.  \label{co4}
\end{eqnarray}
>From expression (\ref{c15}) it follows that the Faddeev-Popov
ghosts $\left( c^{a},\overline{c}^{a}\right) $ and the $i$-valued fields $%
\left( \varphi _{i}^{a},\omega _{i}^{a},\overline{\varphi }_{i}^{a},%
\overline{\omega }_{i}^{a}\right) $ have a common renormalization
constant, given by
\begin{equation}
Z_{c}=Z_{\overline{c}}=Z_{\varphi }=Z_{\overline{\varphi }}=Z_{\omega }=Z_{%
\overline{\omega }}=\left( 1-\eta a_{2}\right)
=Z_{g}^{-1}Z_{A}^{-1/2}\;.  \label{co5}
\end{equation}
Eq.(\ref{co5}) expresses a well-known
renormalization property of the Faddeev-Popov ghosts $\left( c^{a},%
\overline{c}^{a}\right) $ in the Landau gauge, stemming from the
transversality of the gauge propagator and from the factorization of
the ghost momentum in the ghost-antighost-gluon vertex. We see
therefore that, in the present case, this property holds for the
$i$-valued fields $\left(
\varphi _{i}^{a},\omega _{i}^{a},\overline{\varphi }_{i}^{a},\overline{%
\omega }_{i}^{a}\right) $ as well. Similarly to the ghost and the
$i$-valued fields, the renormalization of the sources $\left( M_{\mu
}^{ai},N_{\mu }^{ai},V_{\mu }^{ai},U_{\mu }^{ai}\right) $ is also
determined by the renormalization constants $Z_{g}$ and
$Z_{A}^{1/2}$, being given by
\begin{equation}
Z_{M}=Z_{N}=Z_{V}=Z_{U}=Z_{g}^{-1/2}Z_{A}^{-1/4}\;.  \label{co6}
\end{equation}
It is worth noticing here that equation (\ref{co6})  ensures that
the counterterm $a_{2}\left( V_{\mu }^{ai}M_{\mu }^{ai}-U_{\mu
}^{ai}N_{\mu }^{ai}\right) $ can be automatically reabsorbed by the term $%
\left( -M_{\mu }^{ai}V_{\mu }^{ai}+U_{\mu }^{ai}N_{\mu }^{ai}\right)
$ in the expression (\ref{r5}) without the need of introducing new
free parameters. Indeed,
\begin{equation}
-M_{o}V_{o}=-MVZ_{M}^{2}=-MVZ_{g}^{-1}Z_{A}^{-1/2}=-MV+\varepsilon
a_{2}MV\;. \label{co7}
\end{equation}
Concerning now the parameters $a_{7}$, $a_{8}$, they are easily seen
to correspond to a multiplicative renormalization of the local
source $\tau $ and of the parameter $\zeta $, according to
\begin{eqnarray}
\tau _{o} &=&Z_{\tau }\tau \;,\;\;\;\;\;Z_{\tau }=1+\eta
(a_{7}-2a_{2}+a_{0})\;,  \nonumber \\
\zeta _{o} &=&Z_{\zeta }\zeta \;,\;\;\;\;\;Z_{\zeta }=1+\eta
(-a_{8}-2a_{7}+4a_{2}-2a_{0})\;.  \label{co8}
\end{eqnarray}
Moreover, we would like to underline that there exists even an extra
relation, namely
\begin{equation}\label{extrarel}
    Z_\tau=Z_gZ_A^{-1/2}\;.
\end{equation}
It can be proven by introducing the operator $A_\mu^2$ through a
more sophisticated set of local sources, like it was done in
\cite{Dudal:2002pq}. We will not repeat that analysis here, we only
mention that a key ingredient in the proof of relation (
\ref{extrarel}) was the presence of the ghost Ward identity, and
since the Zwanziger action possesses that identity, eq.(\ref{r13}),
one can proceed along the lines of \cite{Dudal:2002pq}. Thus, there
are in fact only three independent renormalization factors present.

\section{Appendix B.}
In this Appendix, we give the detailed analysis of the procedure
used to optimize the renormalization scheme and scale dependence,
which was summarized in section 5.

\subsection{Preliminaries.} Before coming to the actual computations,
let us first discuss some results which will turn out to be
useful.\\\\Consider again the action $S$ of eq.(\ref{f1}). Due to
the rich symmetry structure of the model, encoded in the Ward
identities (\ref{r9})-(\ref{r22}), and due to the extra relation
(\ref{extrarel}), only three renormalization factors remain to be
fixed, namely $Z_g$, $Z_A$ and $Z_\zeta$. Apparently, this means
that we would need three renormalization conditions in order to fix
a particular renormalization scheme. However, taking a look at the
bare action associated with expression eq.(\ref{f1}), we would find
the following relations
\begin{eqnarray}\label{new17}
\zeta_o&=&Z_\zeta\zeta\;,\nonumber\\
\zeta_o\tau_o^2&=&\omu^{-\varepsilon}Z_\zeta\zeta\tau^2\;,\nonumber\\
\tau_o&=&Z_\tau\tau\;,
\end{eqnarray}
from which it follows that
\begin{equation}\label{new18}
    Z_\zeta\zeta=\omu^{\varepsilon}\zeta_oZ_\tau^2\;.
\end{equation}
Since the bare quantity $\zeta_o$ is renormalization scheme and
scale independent and since $\zeta$ always appears in the
combination $Z_\zeta\zeta$ in the action, it follows that only $Z_g$
and $Z_A$ are relevant for the effective action, because $Z_\tau$
can be expressed in terms of these two factors. Consequently, we
would only need two renormalization conditions to fix the scheme.
Obviously, we can equally well choose to make use of, for example,
$Z_g$ and $Z_\tau$ as the two independent renormalization factors,
corresponding to coupling constant and mass renormalization.
\\\\We will change from the $\MSbar$ to another massless
renormalization scheme by means of the following
transformations\footnote{Barred quantities refer to the $\MSbar$
scheme.}
\begin{eqnarray}\label{trans}
\og^2&=&g^2\left(1+b_0g^2+b_1g^4+\cdots\right)\;,\nonumber\\
\ol&=&\lambda\left(1+c_0g^2+c_1g^4+\cdots\right)\;,\nonumber\\
\om^2&=&m^2\left(1+d_0g^2+d_1g^4+\cdots\right)\;,
\end{eqnarray}
where the parameters $b_i$, $c_i$ and $d_i$ label the new scheme.
However, we should keep in mind that the renormalization of the
Gribov parameter $\lambda$ is not independent of that of $g^2$ and
$m^2$. Eliminating $\gamma_A(g^2)$ between eqns.(\ref{new19}) and
(\ref{new20}), yields
\begin{equation}\label{new21}
\gamma_\lambda(g^2)=\frac{1}{4}\left(\frac{\beta(g^2)}{g^2}-\gamma_{m^2}(g^2)\right)\;.
\end{equation}
This relation, valid to all orders of perturbation theory, implies
the existence of relationships between the coefficients $b_i$, $c_i$
and $d_i$. For further use, we shall explicitly construct the
relation between $b_0$, $c_0$ and $d_0$. Let us adopt as
parametrization of $\beta(g^2)$, $\gamma_{m^2}(g^2)$ and
$\gamma_\lambda(g^2)$
\begin{eqnarray}\label{new22}
\beta(g^2)&=&-2\left(\beta_0g^4+\beta_1g^6+\cdots\right)\;,\nonumber\\
\gamma_{m^2}(g^2)&=&\gamma_0g^2+\gamma_1g^4+\cdots\;,\nonumber\\
\gamma_\lambda(g^2)&=&\lambda_0g^2+\lambda_1g^4+\cdots\;,
\end{eqnarray}
and an analogous one in the case of the $\MSbar$ scheme. Then, one
computes
\begin{eqnarray}\label{new23}
\omu\frac{\p\ol}{\p\omu}&=&\omu\frac{\p}{\p\omu}\left[\lambda\left(1+c_0g^2+\cdots\right)\right]\nonumber\\
&=&\cdots\nonumber\\
&=&
\lambda\left(\lambda_0g^2+\left(\lambda_1+c_0\lambda_0-2\beta_0c_0\right)g^4+\cdots\right)\;,
\end{eqnarray}
which can be expressed in terms of $\gamma_i$ and $\beta_i$ by
exploiting the relation (\ref{new21}). We find
\begin{eqnarray}\label{new24}
\omu\frac{\p\ol}{\p\omu}=
\lambda\left[\frac{-2\beta_0-\gamma_0}{4}g^2+\left(\frac{-2\beta_1-\gamma_1}{4}+c_0\frac{-2\beta_0-\gamma_0}{4}-2\beta_0c_0\right)g^4+\cdots\right]\;.
\end{eqnarray}
We can also calculate $\omu\frac{d\ol}{d\omu}$ by first exploiting
the relation (\ref{new21}), obtaining
\begin{eqnarray}\label{new25}
\omu\frac{\p\ol}{\p\omu}&=&\frac{1}{4}\left[(-2\beta_0-\gamma_0)\og^2+(-2\beta_1-\overline{\gamma}_1)\og^4+\cdots\right]\left[\lambda\left(1+c_0g^2+\cdots\right)\right]\nonumber\\
&=&\cdots\nonumber\\
&=&\frac{1}{4}\left[(-2\beta_0-\gamma_0)g^2+\left(c_0(-2\beta_0-\gamma_0)-2\beta_1-\gamma_1-2\beta_0(-d_0+b_0)\right)g^4+\cdots\right]\;.
\end{eqnarray}
In the previous expression, we had to express $\overline{\gamma}_1$
in terms of $\gamma_1$; a task accomplished by using the relation
\begin{equation}\label{new26}
    \overline{\gamma}_1=\gamma_1-2\beta_0d_0-\gamma_0b_0\;,
\end{equation}
which can be obtained along the same lines of the previous
calculations. It should also be noted that $\gamma_0$, $\beta_0$ and
$\beta_1$ are renormalization scheme independent quantities. Thus,
the identification of eqns.(\ref{new24}) and (\ref{new25}) gives the
desired relation, given by
\begin{equation}\label{new27}
    c_0=\frac{1}{4}(b_0-d_0)\;.
\end{equation}
We now perform the transformations (\ref{trans}) on the action
(\ref{new5}), which was calculated in the $\MSbar$ scheme, to obtain
it in a general scheme.
\begin{eqnarray}\label{new30}
\Gamma  &=&- \frac{\left( N^{2}-1\right)\lambda
^{4}}{2g^2N}\left(1+4c_0g^2-b_0g^2\right)+\frac{\zeta_0m^4}{2g^2}\left(
1-\frac{\zeta _{1}}{\zeta _{0}}g^{2}+2d_0g^2-b_0g^2\right)
\nonumber\\&+&\frac{3\left( N^{2}-1\right)
}{256\pi^2}\left[\left(m^2+\sqrt{m^4-\lambda^4}\right)^2
\left(\ln\frac{m^2+\sqrt{m^4-\lambda^4}}{2\omu^2}-\frac{5}{6}\right)\right.\nonumber\\
&+&\left.\left(m^2-\sqrt{m^4-\lambda^4}\right)^2
\left(\ln\frac{m^2-\sqrt{m^4-\lambda^4}}{2\omu^2}-\frac{5}{6}\right)\right]\;,
\end{eqnarray}
while the gap equations now read
\begin{eqnarray}
\label{new31a}
\frac{\p\Gamma}{\p\lambda}&=&-\frac{2\left(N^2-1\right)}{g^2N}\lambda^3\left(1+4c_0g^2-b_0g^2\right)+\frac{3\left(N^2-1\right)\lambda^3}{256\pi^2}
\left[\frac{8}{3}\vphantom{-4\frac{\left(m^2+\sqrt{m^4-\lambda^4}\right)}{\sqrt{m^4-\lambda^4}}\ln\frac{m^2+\sqrt{m^4-\lambda^4}}{2\omu^2}}\right.\nonumber\\&&\left.-4\frac{\left(m^2+\sqrt{m^4-\lambda^4}\right)}{\sqrt{m^4-\lambda^4}}\ln\frac{m^2+\sqrt{m^4-\lambda^4}}{2\omu^2}+4\frac{\left(m^2-\sqrt{m^4-\lambda^4}\right)}{\sqrt{m^4-\lambda^4}}\ln\frac{m^2-\sqrt{m^4-\lambda^4}}{2\omu^2}\right]\;,\nonumber
\end{eqnarray}
\begin{eqnarray}
\label{new31b} \frac{\p\Gamma}{\p m^2}&=&
\frac{\zeta_0m^2}{g^2}\left(1-\frac{\zeta_1}{\zeta_0}g^2+2d_0g^2-b_0g^2\right)\nonumber\\&+&\frac{3\left(N^2-1\right)}{256\pi^2}
\left[2\left(m^2+\sqrt{m^4-\lambda^4}\right)\left(1+\frac{m^2}{\sqrt{m^4-\lambda^4}}\right)\ln\frac{m^2+\sqrt{m^4-\lambda^4}}{2\omu^2}\right.\nonumber\\
&+&\left.2\left(m^2-\sqrt{m^4-\lambda^4}\right)\left(1-\frac{m^2}{\sqrt{m^4-\lambda^4}}\right)\ln\frac{m^2-\sqrt{m^4-\lambda^4}}{2\omu^2}-\frac{8}{3}m^2\right]\;.\nonumber\\
\end{eqnarray}
We mention that, in the case in which $m^2\geq0$, similar algebraic
manipulations as those leading to the condition (\ref{new16}), give
a more general equation
\begin{equation}\label{new32}
    \frac{68}{39}\left(\frac{16\pi^2}{g^2N}\right)+\frac{161}{39}+\frac{16\pi^2}{N}\left(\frac{32}{3}c_0-\frac{68}{39}b_0-\frac{24}{13}d_0\right)=\frac{t}{\sqrt{1-t}}\ln\frac{t}{\left(1+\sqrt{1-t}\right)^2}\;,
\end{equation}
or, using the relation (\ref{new27}),
\begin{equation}\label{new33}
    \frac{68}{39}\left(\frac{16\pi^2}{g^2N}\right)+\frac{161}{39}+\frac{16\pi^2}{N}\left(\frac{12}{13}b_0-\frac{176}{39}d_0\right)=\frac{t}{\sqrt{1-t}}\ln\frac{t}{\left(1+\sqrt{1-t}\right)^2}\;.
\end{equation}
>From this expression, it is apparent that a sensible solution with
$m^2>0$ might exist, depending on the values of the renormalization
parameters $d_0$ ($\sim$ mass renormalization) and $b_0$ ($\sim$
coupling constant renormalization).
\\\\Frequently used are the so-called physical renormalization
schemes whereby, loosely speaking, one demands that the quantum
corrected quantities reduce to the tree level values at a certain
scale $\omu$. However, it turns out that such an approach is not
particularly useful to implement in the current case due to the
presence of the several scales. Therefore, the question arises how
one can make a somewhat motivated choice for the arbitrary
parameters, labeling a certain renormalization scheme. In the next
subsection we shall discuss a way to reduce the freedom in the
choice of the renormalization parameters. The method relies on the
possibility of performing an optimization of the renormalization
scheme dependence, as illustrated in
\cite{VanAcoleyen:2001gf,Dudal:2002zn}.
\subsection{Optimization of the renormalization scheme.}
Consider a quantity $\varrho$ that runs according to
\begin{equation}\label{opt1}
\omu\frac{d\varrho}{d\omu}=\gamma_\varrho(g^2)\varrho\;,
\end{equation}
where
\begin{equation}\label{opt1bis}
\gamma_\varrho(g^2)=\gamma_{\varrho,0}g^2+\gamma_{\varrho,1}g^4+\cdots\;.
\end{equation}
To $\varrho$, we can associate a quantity $\widehat{\varrho}$ that
does not depend on the choice of the renormalization scheme and
which is scale independent. It is defined as
\begin{equation}\label{opt2}
    \wrho = \mf_\varrho(g^2)\varrho\;,
\end{equation}
whereby
\begin{equation}\label{opt3}
    \omu\frac{d}{d\omu}\mf_\varrho(g^2)=-\gamma_\varrho(g^2)\mf_\varrho(g^2)\;.
\end{equation}
It is apparent that $\wrho$ will not depend on the scale $\omu$. It
can also be checked \cite{VanAcoleyen:2001gf,Dudal:2002zn} that
$\widehat{\varrho}$ is left unmodified by a change of the
renormalization scheme, implemented through transformations
analogous to those of eqns.(\ref{trans}). The equation (\ref{opt3})
can be solved in a series expansion in $g^2$ by noticing that
\begin{equation}\label{opt4}
    \omu\frac{d}{d\omu}\mf_\varrho(g^2)\equiv\beta(g^2)\frac{d}{dg^2}\mf_\varrho(g^2)\;.
\end{equation}
Then, the above differential equation can be solved in a series
expansion in $g^2$, more precisely by
\begin{equation}\label{opt5}
    \mf_\varrho(g^2)=(g^2)^{\frac{\gamma_{\varrho,0}}{2\beta_0}}\left(1+\frac{1}{2}
    \left(\frac{\gamma_{\varrho,1}}{\beta_0}-\frac{\beta_1\gamma_{\varrho,0}}{\beta_0^2}\right)g^2+\cdots\right)\;.
\end{equation}
Consider once more the $\MSbar$ action $\Gamma$ given in
eq.(\ref{new5}). We shall now replace the $\MSbar$ variables $\om^2$
and $\ol$ by their renormalization scheme and scale independent
counterparts $\wm^2$ and $\wl$, which are obtained as before. By
inverting eq.(\ref{opt5}), one has
\begin{eqnarray}
  \label{opt6} \om^2 &=& (\og^2)^{-\frac{\gamma_{0}}{2\beta_0}}\left(1-\frac{1}{2}
    \left(\frac{\overline{\gamma}_{1}}{\beta_0}-\frac{\beta_1\gamma_{0}}{\beta_0^2}\right)\og^2+\cdots\right)\wm^2\;, \\
  \label{opt6bis} \ol&=& (\og^2)^{-\frac{\lambda_{0}}{2\beta_0}}\left(1-\frac{1}{2}
    \left(\frac{\overline{\lambda}_{1}}{\beta_0}-\frac{\beta_1\lambda_{0}}{\beta_0^2}\right)\og^2+\cdots\right)\wl\;.
\end{eqnarray}
Moreover, introducing the notations
\begin{eqnarray}\label{opt6tris}
  a &=& -\frac{\gamma_{0}}{2\beta_0}\;,\;\;\;\;\;\;\;\;\;\;\;\;\;\;\;\;\;\;\;\;\;\;\;\;b=-\frac{\lambda_{0}}{\beta_0}\;, \\
  A &=& -\left(\frac{\overline{\gamma}_{1}}{\beta_0}-\frac{\beta_1\gamma_{0}}{\beta_0^2}\right)\;,\;\;\;\;\;\;\;
  B=-2\left(\frac{\overline{\lambda}_{1}}{\beta_0}-\frac{\beta_1\lambda_{0}}{\beta_0^2}\right)\;,
\end{eqnarray}
the one-loop action is rewritten as
\begin{eqnarray}\label{opt7}
\Gamma  &=&- \frac{\left( N^{2}-1\right)}{2N}(\og^2)^{2b}\wl
^{4}\left(\frac{1}{\og^2}+B\right)+\frac{\zeta_0}{2}\wm^4(\og)^{2a}\left(
\frac{1}{\og^2}+A-\frac{\zeta _{1}}{\zeta _{0}}\right)+\frac{3\left(
N^{2}-1\right) }{256\pi^2}\times
\nonumber\\&&\left[\left(\wm^2(\og^2)^a+\sqrt{\wm^4(\og^2)^{2a}-\wl^4(\og^2)^{2b}}\right)^2
\left(\ln\frac{\wm^2(\og^2)^{a}+\sqrt{\wm^4(\og^2)^{2a}-\wl^4(\og^2)^{2b}}}{2\omu^2}-\frac{5}{6}\right)\right.\nonumber\\
&+&\left.\left(\wm^2(\og^2)^{a}-\sqrt{\wm^4(\og^2)^{2a}-\wl^4(\og^2)^{2b}}\right)^2
\left(\ln\frac{\wm^2(\og^2)^{a}-\sqrt{\wm^4(\og^2)^{2a}-\wl^4(\og^2)^{2b}}}{2\omu^2}-\frac{5}{6}\right)\right]\;.\nonumber\\
\end{eqnarray}
The action (\ref{opt7}) is still written in terms of the $\MSbar$
coupling $\og^2$. Performing the first transformation of
(\ref{trans}), $\Gamma$ can be reexpressed as
\begin{eqnarray}\label{opt8}
\Gamma  &=&- \frac{\left( N^{2}-1\right)}{2N}(g^2)^{2b}\wl
^{4}\left(\frac{1}{g^2}+B-b_0+2bb_0\right)\nonumber\\&+&\frac{\zeta_0}{2}\wm^4(g^2)^{2a}\left(
\frac{1}{g^2}+A-b_0+2ab_0-\frac{\zeta _{1}}{\zeta
_{0}}\right)+\frac{3\left( N^{2}-1\right) }{256\pi^2}\times
\nonumber\\&&\left[\left(\wm^2(g^2)^a+\sqrt{\wm^4(g^2)^{2a}-\wl^4(g^2)^{2b}}\right)^2
\left(\ln\frac{\wm^2(g^2)^{a}+\sqrt{\wm^4(g^2)^{2a}-\wl^4(g^2)^{2b}}}{2\omu^2}-\frac{5}{6}\right)\right.\nonumber\\
&+&\left.\left(\wm^2(g^2)^{a}-\sqrt{\wm^4(g^2)^{2a}-\wl^4(g^2)^{2b}}\right)^2
\left(\ln\frac{\wm^2(g^2)^{a}-\sqrt{\wm^4(g^2)^{2a}-\wl^4(g^2)^{2b}}}{2\omu^2}-\frac{5}{6}\right)\right]\;.\nonumber\\
\end{eqnarray}
So far, we have constructed an action which is written in terms of
renormalization scale and scheme independent variables $\wl$ and
$\wm^2$ and the coupling constant $g^2(\omu)$. This is a certain
improvement, since we are not faced anymore with a choice of the
parameters $d_i$, related to the renormalization of the Gribov and
mass parameter. The remaining freedom in the choice of the
renormalization scheme resides in the coupling constant, labeled by
the parameters $b_0, b_1, \ldots$, and in the scale $\omu$.  Of
course, the higher order coefficients $b_i,\, i=1, \ldots$ do not
show up here, since we have restricted ourselves to the one-loop
level. Nevertheless, we will perform one more step, since the
dependence on the coupling constant renormalization can be reduced
to solely $b_0$, by expanding the perturbative series in inverse
powers of
\begin{equation}\label{opt9}
x\equiv\beta_0\ln\frac{\omu^2}{\Lambda^2}\;,
\end{equation}
rather than in terms of $g^2$. For another illustration of this, see
e.g. \cite{VanAcoleyen:2001gf,Dudal:2002zn}. The coupling constant
$g^2$ can be replaced by $x$ since $g^2$ is explicitly determined by
\begin{equation}\label{opt10}
    g^2=\frac{1}{x}\left(1-\frac{\beta_1}{\beta_0}\frac{\ln\frac{x}{\beta_0}}{x}+\cdots\right)\;.
\end{equation}
In \cite{Celmaster:1979km}, the relation between the scale parameter
$\Lambda$, corresponding to a certain coupling constant
renormalization, and that of the $\MSbar$ scheme, $\lms$, was found
to be
\begin{equation}\label{opt11}
    \Lambda= e^{-\frac{b_0}{2\beta_0}}\lms\;.
\end{equation}
One finally gets the expression (\ref{opt12}). We notice that this
alternative expansion is correct up to order
$\left(\frac{1}{x}\right)^0$.\\\\
In principle, we can solve the two equations
(\ref{opt13})-(\ref{opt13bis}) for the two quantities $\wm_*$ and
$\wl_*$, which will be functions of the two remaining parameters
$\omu$ and $b_0$. However, by construction, we know that $\wm$ as
well as $\wl$ should be independent of the renormalization scale and
scheme order by order. This gives us an interesting way to fix these
parameters by demanding that the solutions $\wm_*(\omu,b_0)$ and
$\wl_*(\omu,b_0)$ depend minimally on $b_0$ and $\omu$. Since this
would give a quite complicated set of equations to solve, we can
make life somewhat easier by reasonably choosing the
scale\footnote{This can be motivated thanks to the scale
independence of the $\widehat{\;\;}$-quantities.} $\omu$ in the gap
equations (\ref{opt13})-(\ref{opt13bis}). In analogy to the choice
for $\omu^2$ done in the previous equation (\ref{choiceomu}), we
shall now set
\begin{equation}\label{num1ima}
\omu^2=
\left|\frac{\wm^2x^{-a}+\sqrt{\wm^4x^{-2a}-\wl^4x^{-2b}}}{2}\right|\;,
\end{equation}
In order to proceed, we still have two quantities at our disposal to
fix the remaining parameter $b_0$. In fact, we can also take the
vacuum energy $E_\mathrm{vac}$ in consideration since, being a
physical quantity, it should depend minimally on the renormalization
scheme and scale. Therefore, we could determine the value for $b_0$
by demanding that
\begin{equation}\label{opt14}
\Upsilon(b_0)\equiv\left|\frac{\p\wl_*^4}{\p
b_0}\right|+\left|\frac{\p\wm_*^4}{\p b_0}\right|+\left|\frac{\p
E_\mathrm{vac}}{\p b_0}\right|\;,
\end{equation}
is minimal w.r.t. the parameter $b_0$. This seems to be a reasonable
candidate. When its dependence on $b_0$ is small, then the
dependence of $\wm$, $\wl$ and $\Evac$ on $b_0$ is necessarily small
too. The ideal situation would be that $\Upsilon$ is zero for a
certain $b_0$. If no such an ideal $b_0$ would exist, we weaken the
condition by requiring that $\Upsilon$ is as small as possible. The
condition (\ref{opt14}) to fix $b_0$ can be considered as some kind
of \emph{principle of minimal sensitivity} \`{a} la Stevenson
\cite{Stevenson:1981vj}. An alternative that is sometimes used is a
\emph{fastest apparent convergence criterion}, where it is demanded
that the quantum corrections are as small as possible compared to
the tree level value. For example, if we denote by $\Gamma^{[0]}$
the action to order $\left(\frac{1}{x}\right)^{-1}$ and by
$\Gamma^{[1]}$ to order $\left(\frac{1}{x}\right)^0$, we could
demand that
\begin{equation}\label{facc}
    \left|\frac{\Gamma^{[1]}-\Gamma^{[0]}}{\Gamma^{[0]}}\right|
\end{equation}
is as small as possible when the parameters fulfill the gap equation
describing the vacuum of the theory.\\\\
Before continuing with explicit calculations, let us just remark
here that the other logarithm, namely
$\ln\frac{\wm^2x^{-a}-\sqrt{\wm^4x^{-2a}-\wl^4x^{-2b}}}{2\omu^2}$,
could become large for a small argument, thus when $\wl^4x^{-2b}$
would be small compared to $\wm^4x^{-2a}$. However, it is harmless
since it appears in the form of $u\ln u$, while we know that
$\left.u\ln u\right|_{u\approx0}\approx0$.

\subsection{Numerical results.}
Let us first give some numerical factors we need. From e.g.
\cite{Gracey:2002yt}, we infer that
\begin{eqnarray}
\label{num2}  \beta_1&=&\frac{34}{3}\left(\frac{N}{16\pi^2}\right)^2
\;,\;\;\;\;\;\;\;\;\;\;\gamma_0 =
-\frac{3}{2}\frac{N}{16\pi^2}\;,\;\;\;\;\;\;\;\;\;\;\gamma_1=-\frac{95}{24}\left(\frac{N}{16\pi^2}\right)^2\;,
\end{eqnarray}
and hence, from the relation (\ref{new21}),
\begin{eqnarray}
  \label{num2tris}\lambda_0=
  -\frac{35}{24}\frac{N}{16\pi^2}\;,\;\;\;\;\;\;\;\;\;\;\lambda_1=-\frac{449}{96}\left(\frac{N}{16\pi^2}\right)^2\;.
\end{eqnarray}
This means that, for any $N$, the quantities $a$ and $b$ in
eq.(\ref{opt6tris}) are found to be
\begin{equation}\label{num3}
    a=\frac{9}{44}\;,\;\;\;\;\;\;\;\;\;\;b=\frac{35}{88}\;.
\end{equation}
It is instructive to consider once more the original
Gribov-Zwanziger model by setting $\wm\equiv0$ and by solving the
gap equation (\ref{opt13}). If $\wl_*$ is a solution of this
equation, then it is not difficult to show that the corresponding
vacuum energy is given by
\begin{equation}\label{num3bis}
    \Evac=\frac{3(N^2-1)}{64\pi^2}\frac{\wl_*^4}{4}\;,
\end{equation}
for any choice of $\omu^2$. Thus, also with the improved
perturbative expansion, the vacuum energy of the original
Gribov-Zwanziger is always nonnegative at the lowest order.\\\\Let
us return to the model we were investigating. We solved the gap
equations stemming from (\ref{opt13})-(\ref{opt13bis}) numerically.
\\\\Let us first search for a possible solution of the gap equation
in the region of space determined by $\wm^4x^{-2a}\geq\wl^4x^{-2b}$.
Taking a look at the action (\ref{opt12}), it might be clear that
the gap equations derived from it will be coupled and hence quite
complicated to solve numerically. From the calculational point of
view, it is useful to introduce new variables, defined by
\begin{eqnarray}\label{coord1}
  \omega_1 &=& \frac{\wm^2x^{-a}+\sqrt{\wm^4x^{-2a}-\wl^4x^{-2b}}}{2}\;, \\
  \omega_2&=&\frac{\wm^2x^{-a}-\sqrt{\wm^4x^{-2a}-\wl^4x^{-2b}}}{2}\;,
\end{eqnarray}
with the inverse transformation
\begin{eqnarray}\label{coord2}
\wm^2 x^{-a}=\omega_1+\omega_2\;,\nonumber\\
\wl^4 x^{-2b}=4\omega_1\omega_2\;.
\end{eqnarray}
This defines a mapping from the space $\wm^4 x^{-2a}\geq\wl^4
x^{-2b}>0$ to $\omega_1\geq\omega_2>0$. One checks that the gap
equations (\ref{opt13})-(\ref{opt13bis}) are equivalent to
\begin{eqnarray}
  \label{coord3}\left(\frac{\omega_1}{\omega_1-\omega_2}\frac{\p}{\p\omega_1}-\frac{\omega_2}{\omega_1-\omega_2}\frac{\p}{\p\omega_2}\right)\Gamma(\omega_1,\omega_2) &=& 0\;, \\
 \label{coord3bis}\left(\frac{1}{\omega_1-\omega_2}\frac{\p}{\p\omega_1}-\frac{1}{\omega_1-\omega_2}\frac{\p}{\p\omega_2}\right)\Gamma(\omega_1,\omega_2)&=&0\;.
\end{eqnarray}
We notice that the case in which $\omega_1$ and $\omega_2$ would
become equal, i.e. $\wm^4 x^{-2a}=\wl^4 x^{-2b}$, should be treated
with some extra care. Let us therefore first assume that
$\omega_1>\omega_2$. Then the two equations
(\ref{coord3})-(\ref{coord3bis}) can be recombined to
\begin{eqnarray}
\label{coord4}\frac{\p}{\p\omega_1}\Gamma &=& 0 \;,\\
\label{coord4bis}\frac{\p}{\p\omega_2}\Gamma &=& 0\;.
\end{eqnarray}
The action $\Gamma(\omega_1,\omega_2)$ is explicitly given by
\begin{eqnarray}\label{coord5} \Gamma  &=&- 2\frac{\left(
N^{2}-1\right)}{N}\mho_1\omega_1\omega_2+\frac{\zeta_0}{2}\mho_2(\omega_1+\omega_2)^2\nonumber\\&+&\frac{3\left(
N^{2}-1\right) }{64\pi^2}\left[\omega_1^2
\left(\ln\frac{\omega_1}{\omu^2}-\frac{5}{6}\right)+\omega_2^2
\left(\ln\frac{\omega_2}{\omu^2}-\frac{5}{6}\right)\right] \;.
\end{eqnarray}
where
\begin{eqnarray}
  \mho_1 &=&  x+B+(1-2b)\left(\frac{\beta_1}{\beta_0}\ln\frac{x}{\beta_0}-b_0\right)\;,\\
  \mho_2 &=& x+A-\frac{\zeta
_{1}}{\zeta
_{0}}+(1-2a)\left(\frac{\beta_1}{\beta_0}\ln\frac{x}{\beta_0}-b_0\right)\;.
\end{eqnarray}
It is not difficult to work out the gap equations
(\ref{coord4})-({\ref{coord4bis}}), being given by
\begin{eqnarray}
  \label{coord6}-2\frac{N^2-1}{N}\mho_1\omega_2+\zeta_0\mho_2(\omega_1+\omega_2)+\frac{3(N^2-1)\omega_1}{32\pi^2}\left(-\frac{1}{3}+\ln\frac{\omega_1}{\omu^2}\right) &=& 0\;, \\
  \label{coord6bis}-2\frac{N^2-1}{N}\mho_1\omega_1+\zeta_0\mho_2(\omega_1+\omega_2)+\frac{3(N^2-1)\omega_2}{32\pi^2}\left(-\frac{1}{3}+\ln\frac{\omega_2}{\omu^2}\right) &=&
  0\;.
\end{eqnarray}
>From the explicit expression of the gap equations and of the action
itself in terms of $\omega_1$ and $\omega_{2}$, the advantages of
using these variables should be obvious, since we can decouple the
two gap equations. Explicitly, since $\omu^2=\omega_1$, one finds
from eq.(\ref{coord6}),
\begin{equation}\label{coord7}
    \omega_2=\frac{\frac{N^2-1}{32\pi^2}-\zeta_0\mho_2}{-2\frac{N^2-1}{N}\mho_1+\zeta_0\mho_2}\omega_1\;,
\end{equation}
which can be substituted in the second gap equation
(\ref{coord6bis}), yielding an equation for $\omega_1$ which does
not contain $\omega_2$ anymore. The nominator of eq.(\ref{coord7})
is different from zero, since filling in the numbers gives
\begin{equation}\label{coord8}
    -2\frac{N^2-1}{N}\mho_1+\zeta_0\mho_2=\frac{N^2-1}{4576}\left(-\frac{975}{\pi^2}-\frac{5984}{N}x\right)\neq0\;.
\end{equation}
where we kept in mind that for a meaningful result,
$x\sim\frac{1}{g^2}$, should be positive. \\\\A numerical
investigation of the gap equation (\ref{coord6bis}) using
eq.(\ref{coord7}) revealed that there are no zeros. We conclude that
there are no solutions with $\wm^4 x^{-2a}> \wl^4x^{-2b}$.\\\\Next,
let us find out if a possible solution with $\wm^4 x^{-2a}=
\wl^4x^{-2b}$ or $\omega_1=\omega_2$ might exist. We explicitly
evaluate the gap equations (\ref{coord3})-(\ref{coord3bis}), where
now $\omu^2=\omega_1$,
\begin{eqnarray}\label{coord9}
  \zeta_0\mho_2(\omega_1+\omega_2) -\frac{N^2-1}{32\pi^2}(\omega_1+\omega_2)-\frac{3(N^2-1)}{32\pi^2}\frac{\omega_2^2}{\omega_1-\omega_2}\ln\frac{\omega_2}{\omega_1}&=&  0\;,\\
  2\frac{N^2-1}{N}\mho_1-\frac{N^2-1}{32\pi^2}-\frac{3(N^2-1)}{32\pi^2}\frac{\omega_2}{\omega_1-\omega_2}\ln\frac{\omega_2}{\omega_1} &=&
  0\,.
\end{eqnarray}
>From the foregoing expressions, we infer that the limit
$\omega_1\rightarrow\omega_2$ exists, giving rise to
\begin{eqnarray}\label{coord10}
  \label{coord10}2\zeta_0\mho_2 -\frac{2(N^2-1)}{32\pi^2}+\frac{3(N^2-1)}{32\pi^2}&=&  0\;,\\
  \label{coord10bis}2\frac{N^2-1}{N}\mho_1-\frac{N^2-1}{32\pi^2}+\frac{3(N^2-1)}{32\pi^2} &=&
  0\;.
\end{eqnarray}
This means that we have two equations to solve for the single
quantity $\omega_1$, which is present in $\mho_1$ and $\mho_2$
through the quantity $x$. It would be an extreme coincidence if
these two different equations, which can be rewritten as
\begin{eqnarray}\label{coord10}
  \label{coord10}\frac{18}{13}\mho_2&=&-\frac{N}{32\pi^2}\;,\\
  \label{coord10bis}\mho_1 &=& -\frac{N}{32\pi^2}\;.
\end{eqnarray}
possess a common solution. That this is not the case can be inferred
from the numerical solutions of both equations (\ref{coord10}) and
(\ref{coord10bis}), shown in Figure 5.
\begin{figure}[h]\label{fig3}
\begin{center}
  \scalebox{1.2}{\includegraphics{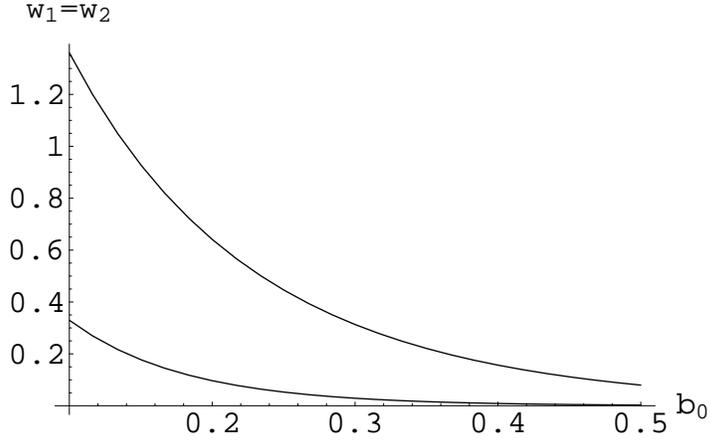}}
\caption{The solution $\omega_1=\omega_2$ as a function of $b_0$ of
eq.(\ref{coord10}), top curve, and eq.(\ref{coord10bis}), bottom
curve, in units $\lms=1$. Clearly, these two curves do no coincide.}
\end{center}
\end{figure}
\\ As a final step, we
should investigate if there is a solution in the region
$\wm^4x^{-2a}<\wl^4 x^{-2b}$. We can still define the coordinates
$\omega_1$ and $\omega_2$ by
\begin{eqnarray}\label{coord11}
  \omega_1 &=& \frac{\wm^2x^{-a}+i\sqrt{-\wm^4x^{-2a}+\wl^4x^{-2b}}}{2}\;, \\
  \omega_2&=&\frac{\wm^2x^{-a}-i\sqrt{-\wm^4x^{-2a}+\wl^4x^{-2b}}}{2}\;.
\end{eqnarray}
In this case, $\omega_1$ and $\omega_2$ are complex conjugate.
Henceforth, it would be more appropriate to use the modulus $R$ and
the argument $\phi$, $\phi\in]-\pi,\pi]$, defined by
\begin{eqnarray}\label{coord12}
  Re^{i\phi} &=& \omega_1\;, \\
  Re^{-i\phi}&=&\omega_2\;,
\end{eqnarray}
If the argument $\phi$ is so that $\left|\phi\right|>\frac{\pi}{2}$,
then $\wm^2 x^{-a}<0$. As a consequence, the estimate for
$\left\langle
A_\mu^2\right\rangle$ will be positive. \\\\
Most of the foregoing analysis can be repeated. The action
(\ref{coord5}) is rewritten in terms of $R$ and $\phi$ by
\begin{eqnarray}\label{coord13} \Gamma  &=&- 2\frac{\left(
N^{2}-1\right)}{N}\mho_1R^2+2\zeta_0\mho_2
R^2\cos^2\phi\nonumber\\&+&\frac{3R^2\left( N^{2}-1\right)
}{32\pi^2}\left[\cos(2\phi)\left(\ln\frac{R}{\omu^2}-\frac{5}{6}\right)-\phi\sin(2\phi)\right]
\;.
\end{eqnarray}
The gap equations (\ref{coord6})-(\ref{coord6bis}) reduce to
\begin{eqnarray}
  \label{coord14}-2\frac{N^2-1}{N}\mho_1R e^{-i\phi}+\zeta_0\mho_2R(e^{i\phi}+e^{-i\phi})+\frac{3(N^2-1)Re^{i\phi}}{32\pi^2}\left(-\frac{1}{3}+i\phi\right) =
  0\;,
\end{eqnarray}
and its complex conjugate. With the parametrization (\ref{coord12}),
we have $\omu^2=R$. \\\\We must solve the following two real
equations\footnote{The $R$-dependence is hidden in $\mho_1$ and
$\mho_2$} for $\phi$ and $R$.
\begin{eqnarray}
  \label{coord15} -2\frac{N^2-1}{N}\mho_1\cos\phi+2\zeta_0\mho_2\cos\phi+\frac{3(N^2-1)}{32\pi^2}\left(-\frac{\cos\phi}{3}
  -\phi\sin\phi\right) &=& 0\;, \\
  \label{coord15bis}2\frac{N^2-1}{N}\mho_1\sin\phi+\frac{3(N^2-1)}{32\pi^2}\left(-\frac{\sin\phi}{3}+\phi\cos\phi\right) &=&
  0\;.
\end{eqnarray}
We can divide these equations\footnote{We may assume
$\cos\phi\neq0$, otherwise eqns.(\ref{coord15})-(\ref{coord15bis})
would give $\phi=0$, which is inconsistent with $\cos\phi=0$.} by
$\cos\phi$ to obtain
\begin{eqnarray}
  \label{coord16} -2\frac{N^2-1}{N}\mho_1+2\zeta_0\mho_2+\frac{3(N^2-1)}{32\pi^2}\left(-\frac{1}{3}-\phi\tan\phi\right) &=& 0 \;,\\
  \label{coord16bis}2\frac{N^2-1}{N}\mho_1\tan\phi+\frac{3(N^2-1)}{32\pi^2}\left(-\frac{\tan\phi}{3}+\phi\right) &=&
  0\;.
\end{eqnarray}
These equations can also be decoupled. The most efficient way to
proceed is to eliminate $R$ between these two equations to obtain an
equation for $\phi$, as the range in we must search for a solution
is limited for this angle. The equation for $\phi$ finally becomes
\begin{equation}\label{oplgap1}
\frac{-90985N-107712\pi^2b_0+12N\left(484\phi\cot\phi+1734\ln\left(\frac{-117\left(50+11\phi\csc\phi\sec\phi\right)}{8228}\right)-1573\phi\tan\phi\right)}{107712\pi^2}=0
\end{equation}
while the value of $R$ is obtained from
\begin{equation}\label{oplgap2}
x\equiv\beta_0\ln\frac{R}{\lms^2}+b_0=-\frac{1950+429\phi\csc\phi\sec\phi}{11968\pi^2}N\;.
\end{equation}
We shall concentrate on the case $N=3$. Depending on the value of
the parameter $b_0$, there is more than one solution possible. In
Figure 6, we have plotted the expression (\ref{oplgap1}) for several
values of the parameter $b_0$, namely
$b_0=0.25,0,-0.25,-0.3,-0.33564...,-0.41594...,-0.5$.
\begin{figure}[h]\label{fig11}
\begin{center}
  \scalebox{1.4}{\includegraphics{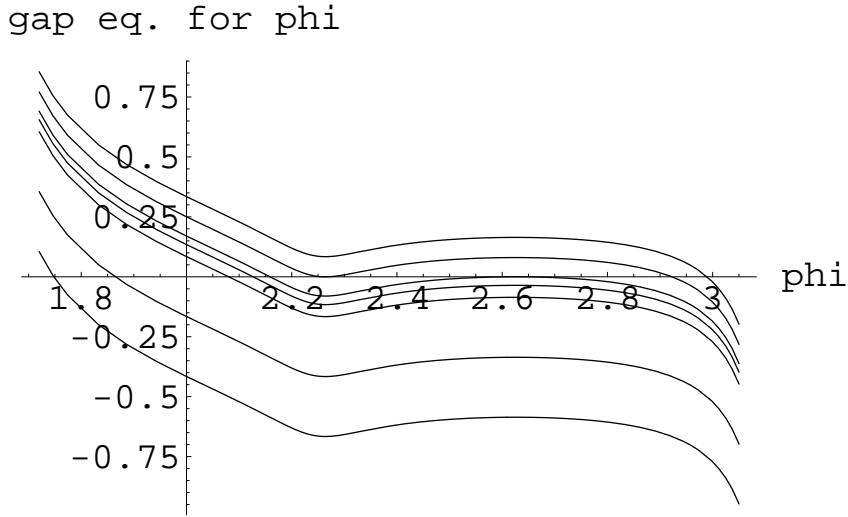}}
\caption{The gap equation (\ref{oplgap1}) with $N=3$ plotted in
function of $\phi$ for the values
$b_0=0.25,0,-0.25,-0.3,-0.33564...,-0.41594...,-0.5$ (from bottom to
top).}
\end{center}
\end{figure}
It is possible to obtain those values of $b_0$ where the number of
solutions change. If we consider the plots of Figure 6, it is
apparent that for each $b_0$, the corresponding curve possesses two
extremal values. The number of solution exactly changes at those
values of $b_0$ where the curve becomes tangent to the $\phi$-axis.
An explicit evaluation learns that his occurs at $b_0=-0.41595...$,
where $\phi=2.26407...$ and at $b_0=-0.33564...$ where
$\phi=2.62545$. It is important to know these numbers to a high
enough accuracy, to instruct the computer in which $\phi$-interval
it can search for a solution. If the initial values are not chosen
in an appropriate way, the iterations will jump between the
different branches of solutions and there will be no convergence to
any of them. There is a single solution $\phi$ if $b_0>-0.33564...$
or $b_0<-0.41595...$. If $-0.41595...< b_0<-0.33564...$, there are
three solutions, while for $b_0=-0.41595...$ and $b_0=-0.33564...$
there are two solutions. In Figure 7, we have displayed the solution
for $\phi$ and $R$.
\begin{figure}[h]\label{fig4and5}
\begin{tabular}{cc}
  \scalebox{0.95}{\includegraphics{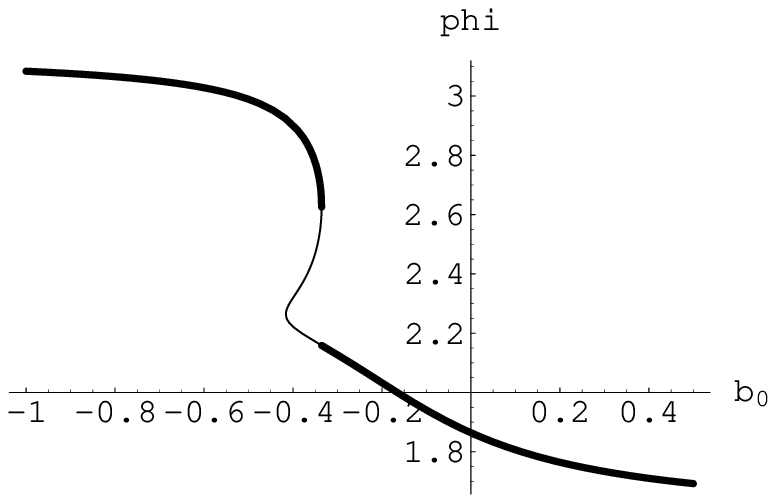}} & \scalebox{0.95}{\includegraphics{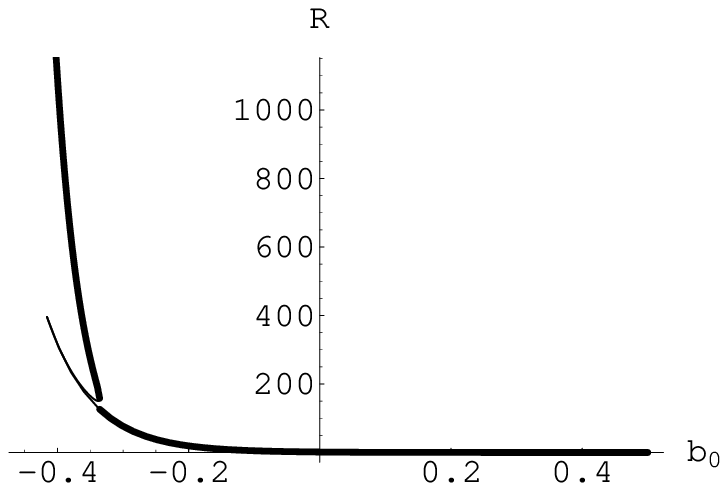}}
  \\
\end{tabular}
\caption{The angle $\phi$ and scale $R$ as a function of $b_0$, in
units $\lms=1$.}
\end{figure}
To determine the solution $\phi$ which characterizes the vacuum, we
should take that one which gives us the absolute minimum of the
energy functional $\Gamma$, which was shown in Figure 3.\\\\
As a final remark, we would like to notice that the same
decomposition as in eq.(\ref{9b}) could also be useful for higher
loop computations. The effective action $\Gamma$ will remain
symmetric under the exchange of $\omega_1$ and $\omega_2$ and
equations like (\ref{coord3})-(\ref{coord4bis}) shall remain valid.
This should facilitate at least a bit the two-loop evaluation of the
effective action and gap equations. Also, one does not need to
evaluate any new anomalous dimension, since these are already known,
either from previous calculations
\cite{Verschelde:2001ia,Browne:2003uv,Dudal:2003by}, or from
exploiting relations like eq.(\ref{new21}).

\end{document}